\documentclass[11pt]{article}
\usepackage{amsmath, amsfonts, amssymb, setspace, url, verbatim, xcolor, cite, graphicx, todonotes, cancel, slashed, mathtools}
\usepackage[T1]{fontenc}
\usepackage{lmodern}
\usepackage[utf8]{inputenc}
\usepackage[
colorlinks=true, % false: boxed links; true: colored links
linkcolor=vub,   % color of internal links
citecolor=vub,   % color of links to bibliography
filecolor=magenta, % color of file links
urlcolor=cyan,
]
{hyperref}
\usepackage[a4paper,top=2cm,left=2.5cm,right=2.5cm,bottom=2cm]{geometry}
\usepackage{microtype}
\newcommand{\be}{\begin{equation}}
\newcommand{\ee}{\end{equation}}
\numberwithin{equation}{section}
\newcommand{\dd}{\text{d}}
\newcommand{\fM}{\mathcal{M}}
\newcommand{\fN}{\mathcal{N}}

\newcommand{\fA}{\mathcal{A}}

\newcommand{\fF}{\mathcal{F}}
\newcommand{\gM}{\mathcal{M}}

\newcommand{\Aa}{\mathcal{A}}
\newcommand{\Ab}{\mathcal{B}}

\newcommand{\Fa}{\mathcal{F}}

\newcommand{\Gfour}{\mathrm{SL}(5)}

\newcommand{\Gsix}{E_{6(6)}}
\newcommand{\Gseven}{E_{7(7)}}

\newcommand{\Edd}{\mathrm{E}_{d(d)}}
\newcommand{\edd}{\mathrm{e}_{d(d)}}

\newcommand{\as}{\mathcal{A}}
\newcommand{\bs}{\mathcal{B}}
\newcommand{\cs}{\mathcal{C}}
\newcommand{\ds}{\mathcal{D}}
\newcommand{\es}{\mathcal{E}}
\newcommand{\fs}{\mathcal{F}}
\newcommand{\gs}{\mathcal{G}}
\newcommand{\hs}{\mathcal{H}}
\newlength{\bibitemsep}
\newlength{\bibparskip}
\let\oldthebibliography\thebibliography
\renewcommand\thebibliography[1]{%
\oldthebibliography{#1}%
\setlength{\parskip}{\bibitemsep}%
\setlength{\itemsep}{\bibparskip}%
}
\definecolor{vub}{RGB}{0,52,154}
\definecolor{vubo}{RGB}{255,102,0}
\definecolor{redd}{RGB}{255,40,40}
\definecolor{r}{RGB}{228,32,20}
\definecolor{o}{RGB}{238,69,4}
\definecolor{y}{RGB}{253,228,1}
\definecolor{g}{RGB}{108,160,0}
\definecolor{b}{RGB}{0,162,203}
\definecolor{i}{RGB}{120,42,117}

\newcommand{\Mint}{\mathbb{M}}

\newcommand{\Ggauging}{G_{\text{gauging}}}

\newcommand{\Gsym}{G_{\text{sym}}}

\usepackage{datetime}

\newcommand{\Mflat}{\bar{M}}
\newcommand{\Aflat}{\bar{A}}
\newcommand{\Bflat}{\bar{B}}
\newcommand{\Fflat}{\bar{F}}
\newcommand{\Df}{\bar{D}}

\newcommand{\Ur}{\mathring{U}}
\newcommand{\gr}{\mathring{g}}
\newcommand{\Mr}{\mathring{M}}
\newcommand{\Urv}{\mathring{\mathbf{U}}}
\newcommand{\Uv}{\mathbf{U}}
\newcommand{\Ev}{\mathbf{E}}
\newcommand{\Xr}{\mathring{X}}

\newcommand{\thetar}{\mathring{\theta}}
\newcommand{\Vr}{\mathring{\mathcal{V}}}
\newcommand{\Er}{\mathring{E}}
\newcommand{\Lambdav}{\boldsymbol{\Lambda}}

\newcommand{\ivec}{I}
\newcommand{\jvec}{J}
\newcommand{\kvec}{K}

\newcommand{\aadj}{a}
\newcommand{\badj}{b}
\newcommand{\cadj}{c}

\begin{document}

\begin{center}

\vspace{-10cm}

\hfill IFT-UAM/CSIC-24-99

\vspace{1em} 

{\LARGE
\bf Infinite and finite consistent truncations \\ 

\vspace{0.3em}
on deformed generalised parallelisations
}

\vspace{1em}
{ \Large Chris D. A. Blair${}^1$, Mart\'in Pico${}^1$, Oscar Varela${}^{1,2,3}$}

\vspace{1em}
{
${}^1$ Instituto de Física Teórica UAM/CSIC, %C/ Nicolás Cabrera 13-15, Universidad Autónoma de Madrid,Cantoblanco, 
Madrid 28049, Spain
\\ ${}^2$ Black Hole Initiative, Harvard University, Cambridge, MA 02138, USA 
\\ ${}^3$ Department of Physics, Utah State University, Logan, UT 84322, USA\\
{\small {\tt c.blair@csic.es}, {\tt martin.pico@uam.es}, {\tt oscar.varela@usu.edu}}
%\\\red{Version \today{} \currenttime}
}

\end{center}

\begin{abstract}
\noindent Given a manifold $\Mint$ admitting a maximally supersymmetric consistent truncation, we show how to formulate new consistent truncations by restricting to a set of Kaluza-Klein modes on $\Mint$ invariant under some subgroup of the group of isometries of $\Mint$.
These truncations may involve either finite or infinite sets of modes.
We provide their global description using exceptional generalised geometry to construct a `deformed' generalised parallelisation starting with that on $\Mint$.
This allows us to explicitly embed known consistent truncations directly into exceptional generalised geometry/exceptional field theory, and to obtain the equations governing situations where the consistent truncation retains an infinite tower of modes.
\end{abstract}

\tableofcontents

\section{Introduction}

A classic problem in (super)gravity is to formulate the dimensional reduction or truncation of a $D$-dimensional theory on some $d$-dimensional manifold, $\Mint$.
This is achieved by keeping some subset of Kaluza-Klein (KK) modes on $\Mint$, such that their interactions can be described solely in $(D-d)$-dimensional terms.
In a consistent truncation, the modes we keep do not source any of the modes we do not, and any lower-dimensional solution of the theory obtained by reduction can be uplifted to a higher-dimensional solution. The simplest example of such a truncation is the standard KK reduction on a circle, $S^1$, keeping only the zero modes of all fields. A more involved example of a consistent truncation is that of 11-dimensional supergravity on the seven sphere, $S^7$. This gives rise to the four-dimensional $\mathrm{SO}(8)$ gauged maximal supergravity \cite{deWit:1986oxb}. 
This retains a particular set of modes on $S^7$. 

In the latter case, $\mathrm{SO}(8)$ corresponds to the isometries of the manifold on which we are reducing.
This paper focuses on the question of whether it is possible to find other consistent truncations by restricting to sets of modes invariant under some (continuous) subgroup $\Gsym \subset \mathrm{SO}(8)$. These truncations will break supersymmetry partially and, depending on the choice and action of $\Gsym$ on the $S^7$, they may involve only a finite set of modes or  infinite towers of modes. The rule of thumb that distinguishes the finite and the infinite cases boils down to the question of whether the action of $\Gsym$ is homogeneous or inhomogeneous. Either way, both will still be truncations (from the set of all possible modes) and will still be consistent (by the usual symmetry considerations). Examples of finite mode truncations include the cases considered in, for example, \cite{Gauntlett:2009zw,Cassani:2011fu}. Here we will describe new infinite-mode truncations.

One powerful explanation for the existence of the maximally-supersymmetric consistent truncation on $S^7$ is provided by adopting the description of supergravity using exceptional generalised geometry (EGG) \cite{Coimbra:2011ky, Coimbra:2012af} or exceptional field theory (ExFT) \cite{Berman:2010is, Hohm:2013pua
}. The maximally supersymmetric consistent truncation on $S^7$ can be viewed to arise from the existence of a generalised parallelisation, i.e., a globally defined nowhere-vanishing frame, of a generalised tangent bundle carrying a representation of $\Gseven$ \cite{Lee:2014mla}. In fact this is the generic exceptional geometric structure underlying maximally-supersymmetric consistent truncations. 
More generally, the EGG/ExFT approaches have proven a very useful framework for constructing and studying consistent truncations with both maximal and less than maximal supersymmetry, in different dimensions e.g. \cite{Hohm:2014qga,Malek:2017njj,Inverso:2017lrz, Cassani:2019vcl,Cassani:2020cod}. 
We will draw our examples from the $S^7$ geometry with the $\Gseven$ EGG/ExFT framework, but our formalism is valid, more generally, for any generalised parallelisable manifold $\Mint$.

In this paper we will show how to start with a consistent truncation arising from a generalised parallelisation and produce from it a new generalised parallelisation characterised by the specification of a symmetry group $\Gsym$, which is a subgroup of the isometries of the original parallelisation manifold.
This new generalised parallelisation can be viewed as a `deformation' of the original one. The original parallelisation will typically be Leibniz, while the deformed one will be non-Leibniz, in the sense that their associated intrinsic torsions will be constant or not. An example of how to retrieve (a subtruncation of) the consistent truncation \cite{Cassani:2011fu} of $D=11$ supergravity on the squashed $S^7$ of \cite{Awada:1982pk} using this type of deformed parallelisations has already appeared in \cite{Duboeuf:2022mam,Duboeuf:2023dmq}. Here we will lay down the systematics of this formalism. 

Our deformed parallelisations also endow the underlying space with a generalised $H$-structure, with $H \subset \Gsym$. For this reason, our work can be also thought of an extension of the formalism of \cite{Cassani:2019vcl,Cassani:2020cod} to the infinite-mode truncation case. For finite consistent truncations, our methods particularise those of \cite{Cassani:2019vcl,Cassani:2020cod} to spaces endowed with an underlying generalised parallelisation. This allows us to obtain full embeddings of the finite truncated theories into EGG/ExFT and make interesting observations. For example, even if the trombone gaugings of the truncated theory vanish, the associated coordinate-dependent trombone in the full ExFT embedding may not.

The formalism we develop is explained in section \ref{sec:formalism}.
In section \ref{sec:ctintro} we review the description of maximally-supersymmetric consistent truncations as generalised parallelisations.
Then in section \ref{logic} we explain how to specify the symmetry group $\Gsym$, construct appropriate singlets under this group and use these to write down a new generalised parallelisation which gives rise to -- infinite and finite -- consistent truncations.
In section \ref{SingletTowers} we connect the (generalised) geometric perspective to the physical picture in terms of towers of KK modes, and illustrate how the geometric and algebraic nature of the singlets defining an infinite consistent truncation can be linked to a (generalised) Kac-Moody algebra.

Then in section \ref{sec:examples} we present explicit examples.
These examples are based on the consistent truncation of 11-dimensional supergravity on $S^7$, whose formulation as a generalised presentation we briefly review in section \ref{GenParallelizationsSevenSphere}.
In section \ref{sec:ExampleSU4}, we discuss the choice $\Gsym = \mathrm{SU}(4)$, and show how this recovers known finite consistent truncations.
In section \ref{sec:ExampleSO7}, we discuss the choice $\Gsym = \mathrm{SO}(7)$.
Here -- due to  $\mathrm{SO}(8)$ triality -- there are three distinct realisations of $\mathrm{SO}(7)$ possible. Two of these are known finite consistent truncations which can in fact be obtained starting with the $\Gsym =\mathrm{SU}(4)$ examples of section \ref{sec:ExampleSU4}.
The third leads to a new infinite consistent truncation, where the fields retain dependence on a coordinate of the internal space, or equivalently an infinite number of modes are kept.

Further discussion is provided in section \ref{sec:discussion}, and diverse technical details are relegated to the appendices.
In particular, in appendix \ref{eom} we present the equations of motion of supergravity in the exceptional field theory formalism in a form adapted to describe finite and infinite consistent truncations.
These equations are used to work out the dynamics of the examples considered in section \ref{sec:examples}.
The remaining appendices \ref{app_7rep}, \ref{appSU4} and \ref{AppendixSO7/SO6Coset} specify group theory details needed for these examples.

%%%%%%%%%%%%%%%%%%%%%%%%%%%%%%%%%%%%%%%%%%%%%%%%%%%%%%%%%%%%%%%%%%%%%%%%
\section{Generalised parallelisations, symmetries and deformations} \label{sec:formalism}

\subsection{Consistent truncations as generalised parallelisations}
\label{sec:ctintro}

We start with a brief overview of how EGG/ExFT \cite{Coimbra:2011ky, Coimbra:2012af, Berman:2010is, Hohm:2013pua}, for more details and references see the review  \cite{Berman:2020tqn}, can be used to describe consistent truncations.
Let us focus on the case of the maximal 11-dimensional supergravity, on some $d$-dimensional manifold $\Mint$. 
Over this manifold we can define various bundles which appear in $\Edd$ generalised geometry.
Generalised vectors are sections of the generalised tangent bundle, $\mathcal{E} \approx T\Mint\oplus \Lambda^2 T^*\Mint \oplus \Lambda^5 T^*\Mint\oplus \dots$.
The fibres of this bundle are isomorphic to a representation $R_1$ of $\Edd$.
Other bundles realise other representations of $\Edd$.

The bosonic symmetries on $\Mint$ are diffeomorphisms generated by $d$-dimensional vectors and gauge transformations generated by $2$- and $5$-forms.
The symmetry transformation parameters combine into a generalised vector $\Lambdav^M$, and their action on the fields realises generalised diffeomorphisms via an action on the generalised Lie derivative, which for instance is defined on a generalised vector of weight $\lambda_V$ by:
\be
\mathcal{L}_{\Lambdav} V^M = \Lambdav^N \partial_N V^M - \alpha \mathbb{P}^M{}_N{}^P{}_Q \partial_P \Lambdav^Q V^N + \lambda_V \partial_N \Lambdav^N V^M \,,
\label{genLie}
\ee
where $\mathbb{P}^M{}_N{}^P{}_Q$ projects into the adjoint of $\Edd$, $\alpha$ is a numerical constant, and we have formally extended the internal coordinates on $\Mint$ into an $\Edd$ representation $y^M$. This extra coordinate dependence is restricted by the so-called section condition, which essentially says that the fields only depend on the actual $d$ coordinates of $\Mint$ (which we will denote by $y^i$), and we will always assume this is the case.
Note as well that we follow the exceptional field theory conventions where generalised vectors which are sections of $\mathcal{E}$ and which generate generalised diffeomorphisms have a non-trivial weight equal to $\tfrac{1}{n-2}$, where $n+d = 11$.

We can always introduce locally a generalised frame field $E^M{}_A$ providing a basis for generalised vectors of weight zero. The inverse $E_M{}^A$ can then be used as a generalised vielbein defining a generalised metric $\gM_{MN} = E_M{}^A E_N{}^B \delta_{AB}$, which parametrises a coset $\Edd/\mathrm{H}_d$, where $\mathrm{H}_d$ is the maximal compact subgroup of $\Edd$.
This generalised metric combines the metric on $\Mint$ together with gauge fields living on $\Mint$.

The other fields of the exceptional field theory approach are the $n$-dimensional external metric, $g_{\mu\nu}$ (which is a scalar of weight $\tfrac{2}{n-2}$ under generalised diffeomorphisms), and a set of gauge fields beginning with a one-form $\Aa_\mu{}^M$ with field strength $\Fa_{\mu\nu}{}^M$.
In this subsection we will only discuss the `internal' sector captured by the generalised metric, and we will extend our discussion to the remaining fields later.

In a conventional consistent truncation to maximal supergravity, the generalised metric admits a factorisation of the form
\be
\gM_{MN}(x,y) = \Ur_M{}^A(y) \Ur_N{}^B(y) \Mr_{AB}(x) \,.
\label{MinCT}
\ee
The matrix $\Ur_M{}^A(y)$ depends only on the coordinates of $\Mint$, and are responsible for selecting the precise modes on $\Mint$ which are kept in the consistent truncation.
Here these are captured in the matrix $\Mr_{AB}(x)$, which provides the scalars of the lower-dimensional theory.
The matrices $\Ur_M{}^A(y)$ should define a \emph{Leibniz generalised parallelisation}.
This means that, introducing a (globally defined) scalar function $\Delta(y)$ of weight $1/(n-2)$ under generalised diffeomorphisms, that 
$\Urv^M{}_A \equiv \Delta \Ur^M{}_A $ provide a globally defined basis for the generalised tangent bundle, and form an algebra 
\be
\mathcal{L}_{\Urv_A}  \Urv_B = - \Xr_{AB}{}^C \Urv_C \,,
\label{UUXU}
\ee
under the $\Edd$ generalised Lie derivative, with \emph{constant} coefficients $\Xr_{AB}{}^C$.
These coefficients can be decomposed in terms of the embedding tensor and trombone gauging (see the general formulae \eqref{defGaugings}): for simplicity we assume here the latter vanishes.

If we decompose the scalar matrix $\Mr_{AB}$ as $\Mr_{AB} = \Vr_A{}^C \Vr_B{}^D \delta_{C D}$, then 
the ansatz \eqref{MinCT} can be viewed as picking a generalised frame of the form $\Er^M{}_A(x,y) = \Ur^M{}_B(y) \Vr_A{}^B(x)$.
Setting $\mathcal{V}_B{}^A = \delta^A_B$, we can see this as a consistent truncation about the background given by $M_{AB} = \delta_{AB}$ (which may or may not be a solution of supergravity).
Then the generalised parallelisation given by $\Er=\Ur \Vr$ can be interpreted as arising from a `deformation' of this background by the scalar modes kept in the consistent truncation.

We want to extend this picture to describe situations where we retain modes that are not present in the consistent truncation on $\Mint$, but retain the structure of generalised parallelisation -- generically with non-constant gaugings appearing in the algebra \eqref{UUXU}.
Effectively, we will do this by making the replacement. \be
\Ur_A (y) \rightarrow P_A{}^B(x,y) \Ur_B (y) \,,
\label{defOutCT}
\ee
with $P_A{}^B \in \Edd$ a globally defined matrix with in principle arbitrary dependence on the coordinates of the internal manifold. 
We will restrict the form of $P_A{}^B$ by requiring that the replacement \eqref{defOutCT} respects some subgroup of the symmetries associated with the consistent truncations \eqref{MinCT}.
This means we will specify some (continuous) symmetry group, $\Gsym \subset \Edd$, and require that $P(y)$ be invariant under the appropriate action of $\Gsym$ described below
Concretely, we will describe an algorithm to obtain (globally defined) $\edd$ Lie algebra-valued functions $S(y)$ which are singlets under the action of $\Gsym$.
Given a set of such singlets we can define $P = \exp(S)$ and use this to construct the deformation \eqref{defOutCT}. 
Because we have imposed the symmetry $\Gsym$, this will still be a consistent truncation, albeit one which may not be a dimensional reduction in the usual sense due to the fact that (as we will describe in detail below) an infinite numbers of modes on $\Mint$ may be retained. This could be viewed as a consistent truncation which is still partially higher-dimensional in nature in that the fields appearing in the ansatz resulting from \eqref{defOutCT} will still depend on some of the coordinates on $\Mint$.

\subsection{Symmetries, singlets, deformations and generalised $G$-structures} 
\label{logic}

\paragraph{Generalised parallelisations and symmetries}

The existence of the Leibniz generalised parallelisation on the manifold $\Mint$ means that we can write down not only a global frame for the generalised tangent bundle, but for all other bundles appearing in the generalised geometry.
Each of these bundles provides particular representations $R_1$, $R_2$, $\dots$ of $\Edd$, and in particular there is always an adjoint bundle.
This can be seen explicitly by noting the usual pairing between tangent and cotangent bundles implies the existence of the generalised covector bundle $\bar{\mathcal{E}} = T^*M \oplus \Lambda^2 TM \oplus \Lambda^5 TM\oplus\dots$ whose fibres are isomorphic to the representation $\bar R_1$ conjugate to $R_1$, and the tensor product $R_1 \otimes \bar R_1$ always contains the adjoint of $\Edd$.
Hence we can speak of Lie algebra-valued functions $f \in C^\infty(\Mint) \otimes \edd$ referring to this global description provided by the parallelisation.

Thanks to this global description, there is a well-defined notion of an $\Edd$ action on such functions. 
However, this $\Edd$ action does not affect $\Mint$ itself. It amounts to internal `rotations' of the generalised parallelisation.

There is however an $\Edd$ action on $\Mint$ via generalised diffeomorphisms.
Consider for simplicity first an $R_1$-valued function, which can be viewed as arising from a set of generalised vectors $f = f^A \Urv_A$.
Under a generalised diffeomorphism with parameter $\Lambdav \equiv \Lambda^A \Urv_A$, with $\Lambda^A$ constant on $\Mint$,
\be
\mathcal{L}_{\Lambdav} f = \Urv_A ( \Lambda^B \partial_B  f^A - \Xr_{BC}{}^A \Lambda^Bf^C ) 
\ee
where $\partial_A \equiv \Urv^M{}_A \partial_M$.
This defines an action specifically of the subgroup $\Ggauging \subset \Edd$ determined by the embedding tensor $\Xr_{AB}{}^C = \Xr_A{}^\alpha t_{\alpha B}{}^C$, thus
\be
\delta_\Lambda f^A = L_v f^A - \Xr_{BC}{}^A \Lambda^B f^C \,.
\ee
where $v$ denotes the vector part of $\Lambdav$.
Here for simplicity we are assuming there is no trombone gauging, so we gauge a subgroup of $\Edd$ (rather than $\Edd \times \mathbb{R}^+$, and accordingly we do not need to be precise about the weights of $f^A$ and $f^\alpha$).
Note that $\Lambda^A$ can be associated with an element of $\Ggauging$ defined by $\Lambda \equiv \Lambda^A \Xr_A{}^\alpha t_\alpha$.
Acting on an adjoint-valued function, whose coefficients are $f^\alpha$, we will similarly have
\be
\delta_\Lambda f^\alpha = L_v f^\alpha - [ \Lambda, f]^\alpha\,,
\label{transf_adj_fn}
\ee
as can be easily verified using the Leibniz rule to obtain the action of the generalised Lie derivative on the adjoint bundle through its action on $\mathcal{E}$ and $\bar{\mathcal{E}}$.
Note that $\Ggauging$ need not necessarily be associated with isometries of $\Mint$.
We can obtain the action of $\Ggauging$ on the metric and form-fields on $\Mint$ by acting with $\Lambdav$ on the generalised metric.
The metric on $\Mint$ appears in the vector-vector components of the generalised metric as $\gM^{ij} = (\det g)^{-1/(n-2)} g^{ij}$, on which the action of $\Lambdav$ reduces to the usual Lie derivative with respect to the vector part of $\Lambdav$. 
We can compute
\be
\mathcal{L}_{\Lambdav} \gM^{MN} = 2 \Ur^M{}_{(A} \Ur^N{}_{B)} \delta_\Lambda \Mr^{AB}\,,\quad
\delta_\Lambda \Mr^{AB} = - 2\Lambda^C \Xr_{CD}{}^{(A}  \Mr^{B)D}\,. 
\ee
Depending on the gauging and the form of $\Mr^{AB}$, this is generically non-zero and we may or may not have Killing vectors.
Let's however assume that we are dealing with special points of the consistent truncation where $\delta_\Lambda \Mr^{AB} = 0$, which guarantees the existence of Killing isometries on $\Mint$.
The main example for us will be the case where $\Mint$ is the round seven-sphere, which defines a consistent truncation to $\mathrm{SO}(8)$ gauged maximal supergravity in four dimensions, with $\Mr_{AB} =\delta_{AB}$.
With this in mind, let's further assume that $\Ggauging$ corresponds to the group of isometries of $\Mint$.

\paragraph{Symmetries and singlets}

Now specify some subgroup $\Gsym \subseteq \Ggauging \subset \Edd$.
Let $a$ be an element of the algebra of $\Gsym$, with the embedding in $\edd$ being given by $a = a{}^A \Xr_A{}^\alpha t_\alpha$.
Our goal now is to characterise all $\edd$-valued singlets invariant under the action of $\Gsym$ induced by the action of generalised diffeomorphisms described above.

Let $S$ denote such a singlet. Using \eqref{transf_adj_fn}, this obeys the defining equation
\be
L_{v[a]} S = [ a, S ]  \,,
\label{SingletEquation}
\ee
where the Killing vector $v[a]$ associated to the Lie algebra element $a$ is defined as the vector part of the generalised vector $a{}^A {\Urv_A}$.

The set of such singlets forms an algebra.
This is easy to verify.
Consider the commutator $[S,S']$, and act on this with $L_{v[a]}$ to get
\be
L_{v[a]} [S,S'] = [ L_{v[a]}S,S'] +  [ S,L_{v[a]}S']
= [ [a,S] , S'] + [ S, [a,S'] ]
= [ a , [S,S'] ] \,,
\ee
using \eqref{SingletEquation} followed by the Jacobi identity. Hence $[S,S']$ obeys the defining equation \eqref{SingletEquation} also.

Now, by assumption there is an action of $\Gsym$ on $\Mint$ as a subgroup of its isometries.
Away from fixed points, this generates orbits $\mathcal{O} \simeq \Gsym / H$ where $H$ is the stabiliser of $\Mint$ under the action of $\Gsym$.
We can show that the algebra of singlets lies within $\mathcal{C}_{\Edd}(H)$ as follows. For any generator $h \in H$ the action of $h$ on functions is given by $ L_{v[h]}$. As $H$ is the stabiliser of the $\Gsym$ action, it follows that there exist a point $p_0 \in \mathcal{O} \subset \Mint$ such that $ L_{v[h]} f(p_0) = 0$ for any $f\in C^\infty(\Mint)$. Then, evaluating \eqref{SingletEquation} at this point we obtain:
\begin{equation}
[ h, S(p_0) ]=0\,,
\end{equation}
which constrains $S$ to be in $\mathcal{C}_{\Edd}(H)$ as a matrix. Thus, the algebra of singlets is contained in the Lie algebra of $\mathcal{C}_{\Edd}(H)$. 
To obtain a solution valid on all of $\Mint$, we view $\Mint$ as a fibration $(\Gsym/H) \hookrightarrow \Mint \rightarrow ( \Mint / ( \Gsym/H) )$.
Then we can solve the singlet equation \eqref{SingletEquation} as
\be
S = L b L^{-1} \,,
\label{SingletSolution} 
\ee
where $L \in \Gsym/H$ parametrises the coset fibres, and $b \in C^\infty(  \Mint / ( \Gsym/H)  ) \otimes \mathcal{C}_{\Edd}(H)$.
 Here, $\mathcal{C}_{G}(H)$ denotes the commutant of $H \subset G$ inside $G$. The latter condition implies that $L_{v[a]} b = 0$ and $h[a]b=bh[a]$. The former condition means that under an infinitesimal $\Gsym$ transformation, $L$ transforms according to the usual coset transformation rule
\be
- L_{v[a]} L  + a L = L h[a]\,,
\ee
where $h[a]$ is a local compensator valued in the Lie algebra of $H$. 
Similarly, we have $-L_{v[a]} L^{-1} - L^{-1} a = - h[a]L$.
A short calculation using the above facts then shows that  \eqref{SingletSolution} obeys \eqref{SingletEquation}.

Note that more generally given a set of generators $b_a$ for $\mathcal{C}_{\Edd}(H)$ we can write $b$ in \eqref{SingletSolution} in the form
\be
b(x,y) = \sum_a \phi^a(x,y) b_a \,,
\ee
where $\phi^a(x,y)$ are arbitrary functions of the coordinates $x$ of the $n$-dimensional external spacetime, while the internal dependence is restricted such that these are functions only on the quotient $\Mint / ( \Gsym/H)$.
In practice, we will be able to choose coordinates $y=(y', \tilde y)$ where $y'$ is a coordinate on $\Gsym/H$ and $\tilde y$ a coordinate on $\Mint/(\Gsym/H)$, such that $L(y) = L(y')$ and $b(y) = b(\tilde y)$. To avoid clutter we will not make this distinction explicitly.
The functions $\phi^a(x,y)$ can be viewed as $y$-dependent extensions of the scalar fields that would appear if we were carrying out a conventional finite consistent truncation.
We can then expand the $y$-dependence in modes on $\Mint$ to view this as a consistent truncation retaining an infinite number of modes. 

\paragraph{Deformations}
The singlets \eqref{SingletSolution} can then be used to construct a new generalised parallelisation, which is a `deformation' of the original one. 
Given some singlet $S(y) = L b L^{-1}$, it exponentiates to an $\Edd$-valued matrix.
We can do this for instance in the $R_1$ representation to define:
\begin{equation}
P_A{}^B(x,y)= \exp (S(x,y))_A{}^B
= ( L(y) \mathcal{V}^{-1}(x,y)L^{-1}(y))_A{}^B\,,
\label{SingletSolutionExponential}
\end{equation}
where we defined $\mathcal{V}^{-1} \equiv \exp\badj $.
We use this matrix \eqref{SingletSolutionExponential} as in \eqref{defOutCT} to define a new generalised parallelisation, such that 
\be
\begin{split} 
\gM_{MN}(x,y) &  = \Ur_M{}^A(y) (P^{-1})_A{}^C(x,y) \Mr_{CD}(x) \Ur_N{}^B(y) (P^{-1})_B{}^D(x,y) 
\\
& = \Ur_M{}^A(y) L(y)_A{}^C \mathcal{V}(x,y){}_C{}^D \Mr_{DE}(x) \mathcal{V}(x,y){}_F{}^E L(y){}_B{}^F \Ur_N{}^B(y) \,,
\end{split}
\label{defGenPar1}
\ee
where in the second line we used the fact that $L^{-1} \in \Gsym \subseteq \Ggauging$ together with our assumption that $\Mr_{AB}$ is preserved by $\Ggauging$.
It follows from our assumptions that the geometry defined by \eqref{defGenPar1} is invariant under generalised diffeomorphisms $\Lambdav = \Lambda^A \Urv_A$ with $\Lambda^A \Xr_A{}^\alpha \in \Gsym$, i.e. the subset of Killing vectors of the original background corresponding to this symmetry group remain Killing vectors of the new solution.

The generalised metric factorisation \eqref{defGenPar1} can be conveniently re-expressed as
\be
\begin{split}
\gM_{MN}(x,y) & = U_M{}^A (y) M_{AB}(x,y) U_N{}^B (y) \,,\\
M_{AB}(x,y) & \equiv \mathcal{V}(x,y){}_A{}^C \Mr_{CD}(x) \mathcal{V}(x,y){}_B{}^D\,,\quad
U_M{}^A  \equiv \Ur_M{}^B L_B{}^A \,.
\end{split}
\label{defMU} 
\ee
This allows us to treat $U_M{}^A$ as defining the new generalised parallelisation.
In terms of $\Uv^M{}_A \equiv \Delta U^M{}_A$ we have 
\be
\mathcal{L}_{\Uv_A} \Uv_B =- X_{AB}{}^C \Uv_C \,,
\ee
defining \emph{non-constant} generalised gaugings $X_{AB}{}^C$.
We can express the resulting embedding tensor and trombone $\Theta$ and $\theta$ evaluated in terms of the generalised parallelisation defined by $U_A$ in terms of those of the original parallelisation $\mathring{U}_A$ as
\be
\begin{split} 
\Theta_{AB}{}^C & = \mathring{T}_{AB}{}^C + \kappa (P_{R_\Theta})_{AB}{}^C,{}^{DE}{}_{F} W_{DE}{}^F \,,\,\quad
\theta_A  =% \mathring{\theta}_B L^B{}_A + 
\tfrac{1}{n-2} \mathring{\partial}_B (L^{-1})_A{}^B \,,\\
\mathring{T}_{AB}{}^C& = (L^{-1})_A{}^D (L^{-1})_B{}^E  \mathring{\Theta}_{DE}{}^F L_F{}^C \,.\\
\end{split}
\label{IntrinsicTorsionComponentsForNonMaximalTruncations} 
\ee
with $W_{AB}{}^C = (L^{-1})_A{}^D (L^{-1})_B{}^E \mathring{\partial}_{D} L_E{}^C$, $\mathring{\partial}_A \equiv \Delta \mathring{U}^M{}_A \partial_M$.
The crucial issue here is that generically $\Theta_{AB}{}^C$ and $\theta_A$  will still be $y$-dependent.
Note as well that even if (as we have been assuming) the original trombone vanishes, $\thetar_A=0$, that $\theta_A$ may be non-zero. 

In addition to the generalised metric, we must specify the truncation ansatz for the $n$-dimensional external metric.
This will be of the form
\be
g_{\mu\nu}(x,y) = \Delta^2(y) \bar g_{\mu\nu}(x,y) \,.
\label{gansatz} 
\ee
This uses the scalar function $\Delta$ inherited unchanged from the original generalised parallelisation we started with.
On the right-hand side, we allow for $\bar g_{\mu\nu}$ to still have an internal coordinate dependence, specifically on the coordinates of $\Mint/(\Gsym/H)$ which likewise appear in $M_{AB}(x,y)$.

\paragraph{Generalised $H$-structures}

We gain further insight into the geometry resulting from \eqref{defGenPar1} by adopting the language of generalised $G$-structures \cite{Cassani:2019vcl}. Firstly, recall that any generalised metric can be decomposed as $\gM_{MN} = E_M{}^A E_N{}^B \delta_{AB}$. 
There is a local $H_d$ invariance under which $E^A$ transforms and $\delta_{AB}$ is invariant.
This defines a reduction of structure from $\Edd$ to $H_d$, i.e. the (generalised) structure group for patching the generalised vielbein is $H_d$.
If $E_M{}^A$ is globally well-defined, giving a generalised parallelisation, then the generalised structure group is trivial.
In between these two extremes we can have other non-trivial generalised structure groups.

The factorisation \eqref{defMU} can be easily seen to be invariant 
under transformations $U_M{}^A \rightarrow N^A{}_B  U_M{}^B$ where $N^A{}_B$ is a local $H$ transformation, using the facts that $\mathcal{V}$ commutes with $H$ in $\Edd$ and the assumption that $H \subseteq \Gsym$ preserves $\Mr_{AB}$.
This implies that the background defined by \eqref{defMU} admits $H$ (or a subgroup thereof) as a generalised structure group.

\paragraph{Gauge field sector} 

We now consider the gauge field sector.
In exceptional field theory, this consists of a hierarchy of external $p$-forms, $\Aa_\mu{}^M$, $\Ab_{\mu\nu \alpha}$, $\dots$, carrying representations $R_p$ of $\Edd$.
With a view to our later applications all in the $\Gseven$ case, we use the index $\alpha$ for the $R_2$ representation, which in that case is indeed the adjoint, and we will not explicitly discuss $p$-forms with $p>2$.

To truly make use of the generalised $H$-structure which we have highlighted above, we should truncate the gauge field sector to consist solely of $H$-invariant fields.
We therefore proceed similarly to \cite{Cassani:2019vcl} and introduce $H$-invariant quantities $K_{\ivec }{}^A\in R_1$ and $J^{\aadj }{}_\alpha \in R_2$ such that for $h$ an arbitrary generator of the Lie algebra of $H \subset \Edd$ we have
\be
h_B{}^A K_{\ivec }{}^B = 0 \,,\quad
h_\alpha{}^\beta J^{\aadj }{}_\beta = 0 \,.
\ee
The index $\ivec $ labels the $H$-invariant generalised vectors, while $\aadj $ labels the $H$-invariant generalised tensors in the $R_2$ representation.
The definition of these quantities is purely algebraic and effected at the level of the global $\Edd$ provided by the generalised parallelisation with respect to $U_A$, so they are genuine constants.
We can then obtain generalised vectors $K_{\ivec } \equiv \Delta U_A K_{\ivec }{}^A$ and tensors $J^{\aadj }{}_\alpha \equiv \Delta^2 U_\alpha{}^\beta J^{\aadj }{}_\beta$ by contracting with the generalised parallelisation -- these correspond to the $K$ and $J$ of \cite{Cassani:2019vcl}.
We can then expand the gauge fields in terms of these objects:
\be
\begin{split} 
\Aa_\mu{}^M(x,y) & = \Delta(y)  U^M{}_A(y) K_{\ivec }{}^A A_\mu{}^{\ivec }(x,y) \,,\\
\Ab_{\mu\nu \alpha}(x,y) & = \Delta^2(y)  U_\alpha{}^\beta(y) J^{\aadj }{}_\beta B_{\mu\nu\aadj }(x,y) \,.
\end{split} 
\label{ABexpand} 
\ee
On the right-hand side we have indicated that the fields $A_\mu{}^{\ivec }$ and $B_{\mu\nu\aadj }$ may still depend on the internal coordinates, specifically on those parametrising $\Mint/(\Gsym/H)$.

The equations of motion for the deformed generalised parallelisation with the commensurate expansion \eqref{ABexpand} follow from the general expressions in appendix \ref{EomGenDef}.

\paragraph{Finite consistent truncations and generalised $H$-structures}
When the intrinsic torsion of a generalised $H$-structure is constant, then it has been established \cite{Cassani:2019vcl} that there is a conventional consistent truncation to a lower-dimensional theory obtained by expanding all fields in terms of the invariant tensors that define the structure.
For the gauge fields, this corresponds to \eqref{ABexpand} with the restriction that $A_\mu{}^{\ivec }$ and $B_{\mu\nu\aadj }$ only depend on $x$ and not the internal coordinates $y$.
Let's further explain how this works in our approach.
We can pick out the singlet torsion by defining
\be
X_{\ivec  A}{}^B \equiv K_{\ivec }{}^C X_{CA}{}^B \,.
\label{singlettorsion}
\ee
This is the quantity that appears in the action of the gauge fields on generalised tensors, e.g. $\mathcal{L}_{\Aa_\mu} V^M = - \Uv^M{}_A A_\mu{}^{\ivec } X_{\ivec  B}{}^A V^B$ for $V^M = \Uv^M{}_A V^A$ a generalised vector.
Note the trombone $\theta_{\ivec } = X_{\ivec A}{}^A$ may be non-zero.

For a consistent truncation, $X_{\ivec A}{}^B$ must lie in $\mathcal{C}_{\Edd}(H)$ and be constant.
The first of these conditions implies that the generalised vectors $K_{\ivec } \equiv K_{\ivec }{}^A \Delta  U_A$ form an algebra
\be
\mathcal{L}_{K_{\ivec }} K_{\jvec } = - X_{\ivec  \jvec  }{}^{\kvec } K_{\kvec } \,,\quad
 X_{\ivec \jvec }{}^{\kvec }  \equiv K_{\ivec }{}^A  K_{\jvec }{}^B X_{AB}{}^C \tilde{K}^{\kvec }{}_C 
 \label{GaugeFieldsGeneralizedDerivativeClosure}
\ee
where $\tilde{K}^{\ivec }{}_A$ denotes a dual set of $H$-invariant quantities such that  $\tilde{K}^{\ivec }{}_A  {K}_{\jvec }{}^A =\delta^{\ivec }_{\jvec }$.

For a standard dimensional reduction, the scalar matrix $M_{AB}$ should be independent of the coordinates on $\Mint$.
The definition \eqref{defMU} shows that $M_{AB}$ depends on these coordinates via $\mathcal{V}_A{}^B$, which lies in $\mathcal{C}_{\Edd}(H)$ meaning that the full scalar coset is in accord with the expectations of \cite{Cassani:2019vcl}.
However, $\mathcal{V}_A{}^B$ in general depends on the coordinates of $\Mint / ( \Gsym/H)$. 
In cases when this quotient turns out to be trivial (in particular for cases where $\Gsym$ acts homogeneously on $\Mint$), then $\mathcal{V}_A{}^B$ will only depend on the external coordinates $x$ and we will be able to use $M_{AB}$ as the scalar matrix for a lower-dimensional theory. 
We then also restrict the external metric ansatz \eqref{gansatz} such that $\bar g_{\mu\nu} = \bar g_{\mu\nu}(x)$ only.
Furthermore, if the generalised $H$-structure branches the $\mathbf{8}_s$ representation leaving $k$ singlets, we expect to recover a supersymmetric truncation with $\mathcal{N}=k$ supersymmetry \cite{Cassani:2019vcl}.

The equations of motion for the dimensionally reduced theory obtained by this finite consistent truncation then follow as in appendix \ref{EomForNonMaximalTruncations}.

Note that additional consistency requirements will appear when considering the field strengths of the gauge fields.
Let's focus for simplicity on the case of most interest to the applications in this paper, which is the $\Gseven$ ExFT \cite{Hohm:2013uia, Hohm:2014qga}.
When we expand the field strength $\Fa_{\mu\nu}{}^M= \Delta U^M{}_A F_{\mu\nu}{}^A$ of this theory using \eqref{ABexpand},\footnote{This field strength involves an additional `constrained compensator' two-form for which the appropriate ansatz will be \cite{Hohm:2014qga} $\Ab_{\mu\nu M}(x,y) = - 2 \Delta^2 (y) U^P{}_B(y) \partial_M U^A{}_P(y) t^\alpha{}_A{}^B J^{\aadj }{}_\alpha B_{\mu\nu \aadj } (x)$.
} we find a very specific combination of fluxes multiplying the two-forms:
\be
F_{\mu\nu}{}^A  = 
K_{\ivec }{}^A ( 
2  \partial_{[\mu}A_{\nu]}{}^{\ivec }
+ X_{\jvec \kvec }{}^{\ivec } A_{[\mu}{}^{\jvec } A_{\nu]}{}^{\kvec })
+ ( X^{A \alpha}  - 24 t^{\alpha AB } \theta_B ) J^{\aadj }{}_{\alpha} B_{\mu\nu \aadj } 
\label{Fpartway}
\ee
Given that $J^{\aadj }{}_{\alpha}$ is valued in $\mathcal{C}_{\Gseven}(H)$, we have to require that the projection of $X^{A \alpha}  - 24 t^{\alpha AB } \theta_B$ to this commutant is $H$-invariant and constant.
Then we obtain $F_{\mu\nu}{}^A = F_{\mu\nu}{}^{\ivec } K_{\ivec }{}^A$ with
\begin{equation}
\begin{split}
 F_{\mu\nu}{}^{\ivec } &= 2  \partial_{[\mu}A_{\nu]}{}^{\ivec } + X_{\jvec \kvec }{}^{\ivec } A_{[\mu}{}^{\jvec } A_{\nu]}{}^{\kvec } \\
&\quad \quad + \Omega^{\ivec \jvec }K_{\jvec }{}^A( X_{AB}{}^{C}  - 24  \mathbb{P}^C{}_B{}^D{}_A \theta_D )J^{\aadj }{}_C{}^B B_{\mu\nu \aadj }\,,
\end{split} 
\label{FieldStrengthsTruncation}
\end{equation}
where $\Omega^{\ivec \jvec } \equiv \tilde{K}^{\ivec }{}_A \Omega^{AB} \tilde{K}^{\jvec }{}_B$ is just the restriction of the $\Gseven$ symplectic invariant to our truncation.

\subsection{Singlets, Kaluza-Klein towers and generalised Kac-Moody algebras}
\label{SingletTowers}

In this subsection, we discuss the connection between the $\Gsym$-singlets $S(y)$ characterised above and the tower of $\Gsym$-invariant KK modes on $\Mint$.
We will show that the formula \eqref{SingletSolution}, expressing the singlets in terms of the coset element $L \in \Gsym/H$ and the $\mathcal{C}_{\Edd}(H)$-valued matrix $b$, can be viewed as capturing a rearrangement of a (possibly infinite) tower of KK modes. 

To discuss this explicitly, we will focus on two instructive examples based on the consistent truncation of 11-dimensional supergravity on the round $S^7$, for which $\Ggauging = \mathrm{SO}(8)$.
This is known to be described using a generalised parallelisation in $\Gseven$ generalised geometry whose details we will review in section \ref{GenParallelizationsSevenSphere}.

For the moment, we just need the basic fact that we can define the sphere $S^7 \subset \mathbb{R}^8$ in terms of embedding coordinates $\mu^\as$, which obey $\delta_{\as \bs} \mu^\as \mu^\bs =1$ and transform as the $\mathbf{8}_v$ of $\mathrm{SO}(8)$.
The spherical harmonics on $S^7$ then consist of symmetric traceless combinations of the $\mu^\as$.
We can organise these into KK levels based on the number of powers of $\mu^\as$ appearing.
Thus at level 0 we have constant functions on $S^7$, at level one we have the embedding coordinates $\mu^\as$ themselves, at level two we have quadratic powers $\mu^\as \mu^\bs$, and so on. 
Any smooth function $f(y)$ on $S^7$ can be expanded in terms of harmonics as $f(y) = f^\Sigma \mathcal{Y}_\Sigma$, where $\mathcal{Y}_\Sigma$ runs on the infinite set of all the $S^7$ harmonics.

Now let's take some $\Gsym \subseteq \mathrm{SO}(8)$. At the Lie algebra level, the $\Gsym$-invariant singlets we need are sections of $C^\infty(\Mint) \otimes \edd$. Any such a singlet $S$, can be expanded in terms of harmonics as $S= S^\Sigma \mathcal{Y}_\Sigma$, where $S^\Sigma = S^{\alpha\Sigma} t_{\alpha} \in \edd$. The adjoint $\alpha$ can be branched under $\mathrm{SO}(8)$, so $S=S^{\alpha\Sigma} \mathcal{Y}_\Sigma t_{\alpha}$ can always be expressed in terms of fundamental $\mathrm{SO}(8)$ indices.
The singlets under $\Gsym \subseteq \mathrm{SO}(8)$ are obtained by further branching $\mathrm{SO}(8)$ under $\Gsym$ and restricting $S^{\alpha\Sigma}$ to be $\Gsym$ invariant tensors. Let us show this with two different examples leading to two different structures of the KK tower of modes.

\paragraph{$\mathrm{SU}(4)_c$ symmetry (homogeneous case).} Under $\mathrm{SU}(4)_c$ we have the following branching of the fundamental representations of $\mathrm{SO}(8)$:
\begin{equation}
\mathbf{8}_v  \rightarrow \mathbf{4} + \mathbf{\bar{4}}\,, \quad
\mathbf{8}_c  \rightarrow \mathbf{6} + \mathbf{1} + \mathbf{1} \,, \quad
\mathbf{8}_s  \rightarrow \mathbf{4} + \mathbf{\bar{4}} \,.
\label{SO8branchSU4}
\end{equation}
We will not need to explicitly introduce fundamental $\mathrm{SU}(4)$ indices.
Instead, to construct singlets we will make use of the existence of the $\mathrm{SU}(4)$ invariant tensors, $J_{\as \bs}$ and $\Omega_{\as \bs \cs \ds}$ which we write carrying $\mathbf{8}_v$ indices.
These invariants are antisymmetric and defined in the appendix in equation \eqref{OmegaJ}.
The real and imaginary parts of the rank-four invariant $\Omega$ gives two real invariants, $\mathrm{Re} \,\Omega$ and $\mathrm{Im}\, \Omega$, each of which is self-dual.

Using invariant tensors we can systematically construct all $\mathrm{SU}(4)_c$ invariant singlets working level by level in the tower of KK modes i.e. the tower of spherical harmonics. 
In terms of the $\Gseven$ generators branched under $\mathrm{SO}(8)$ as in \eqref{Rstpm}, where $R_{\as \bs} \in \mathbf{28}$ is antisymmetric, $S_{\as \bs} \in \mathbf{35}_v$ is symmetric and $t^\pm_{\as\bs\cs\ds} \in \mathbf{35}_{c/s}$ are antisymmetric and self-dual/anti-self-dual, we find the following.
\begin{itemize}
\item Level 0. There are four independent singlets involving no $\mu^\as$:
\begin{equation}
\begin{array}{ll}
S_1 = \text{Re}\, \Omega^{\as \bs \cs \ds} \;t^+_{\as \bs \cs \ds} \,,  &S_2 = \text{Im}\, \Omega^{\as \bs \cs \ds} \;t^+_{\as \bs \cs \ds} \,,\\
S_3 =  \tfrac12 J^{\as \bs} \;R_{\as \bs} \,, &S_4 = J^{\as \bs} J^{\cs \ds} \;t^+_{\as \bs \cs \ds}\,.
\end{array}
\end{equation}
\item Level 1. There are no possible singlets involving a single $\mu^\as$.
\item Level 2. There are seven independent singlets involving two $\mu^\as$:
\begin{equation}
\begin{matrix}
\begin{array}{ll}
S_5 =\tfrac12 \mu^\as \mu^\bs \;S_{\as \bs}  \,,  &S_6 = \tfrac12 \mu^\as \mu_\ds  J^{\cs \ds} \;S_{\as \cs} \,,\\
S_7 =  \tfrac12 \mu_\cs \mu_\ds  J^{\as\cs } J^{\bs \ds} \;S_{\as \bs}  \,, &S_8 = \text{Re}\, \Omega^{\as \bs \cs \es} \mu_\es \mu^\fs \;t^-_{\as \bs \cs \fs} \,,\\
S_9= \text{Im}\, \Omega^{\as \bs \cs \es} \mu_\es \mu^\fs \;t^-_{\as \bs \cs \fs} \,, & S_{10}= J^{\as\bs } J^{\cs\ds } \mu_\ds \mu^\es \;t^-_{\as \bs \cs \es}\,,
\end{array}\\
S_{11}= \tfrac12 \mu^\as \mu_\cs  J^{\bs \cs} \;R_{\as \bs} \,.
\end{matrix}
\end{equation}
\item Higher levels. There are no possible singlets involving more than two $\mu^\as$.
\end{itemize} 
We therefore find a finite number of singlets.
Taking appropriate linear combinations, we find that the singlets span the algebra $\mathrm{SL}(2) \times \mathrm{SU}(2,1)$ with the same commutation relations as the one appearing in appendix \ref{appSU4}. The correspondence is made as follows:
\begin{equation}
\begin{array}{llll}
H_0 = 2 (S_5 + S_7)\,, &  E_0 = S_4 + 8 S_{10} \,, & F_0 = \tfrac{9}{64} (S_4 - 8 S_{10})\,, & \\
H_1=S_5 -S_7\,, & H_2 = S_3 +2S_{11} \,, & E_2 = 128 \sqrt{2} (S_6 + S_{11})\,, & F_2 = \tfrac{1}{64 \sqrt{2}} (S_6 - S_{11})\\
E_{11} = S_1 + 8 S_8\,, & E_{12} = S_2 + 8 S_9\,, & F_{11} =\tfrac{1}{128} ( S_1 - 8 S_8)\,, & F_{12} =\frac{1}{128}( S_2 - 8 S_9)\,.
\end{array}
\end{equation}
Now, recall that our geometrically obtained singlets \eqref{SingletSolution} were of the form $S = L b L^{-1}$.
This expression can be viewed as a similarity transform using the matrix $L \in \mathrm{SU}(4) / H$ acting on $b$. The latter was algebraically was an element of the commutant of $H$, where $H$ arose as the stabiliser of the $\mathrm{SU}(4)_c$ action on $S^7$.
Indeed in this case it can be verified that the stabiliser of $S^7$ under the action of $\mathrm{SU}(4)_c$ is $\mathrm{SU}(3)$, as $S^7 \cong \mathrm{SU}(4)_c/\mathrm{SU}(3)$, and furthermore $\mathrm{SL}(2) \times \mathrm{SU}(2,1) = \mathcal{C}_{\Gseven}\left(\mathrm{SU}(3)\right)$.
Noting that there is a unique coordinate-independent realisation of $\mathcal{C}_{\Gseven}\left(\mathrm{SU}(3)\right)$ within $E_{7(7)}$ (see appendix \ref{appSU4}), it follows that there must exist a coordinate-dependent matrix $L(y) = L(\mu(y))$ such that both bases are related. 
Thus, for any coordinate-dependent $\tilde{b}(y)$ in $\mathcal{C}_{\Gseven}\left(\mathrm{SU}(3)\right)$ we have $\tilde{b}(y) = L(y) b L(y)^{-1} $, where $b$ is coordinate-independent.

In addition, from the general formula \eqref{SingletSolution} for singlets we note that in this case  $\Mint / (\Gsym/H)$ is trivial and hence the singlets $\tilde{b}(y)$ do not have arbitrary coordinate dependence on this quotient. This implies that we must find a finite number of singlets.

We therefore see that the role of $L(y)$ in this case is to rotate the algebra to a basis where the algebra of singlets can be expressed in terms of a coordinate-independent algebra. It is also worth remarking that the $L(y)$ provided in the appendix \ref{appSU4} in equation \eqref{SU4cL(y)56from8v} can be expressed in terms of level 2 KK modes, in correspondence with the fact that the highest singlets in the tower for $\mathrm{SU}(4)_c$ lie at level 2. 

This example corresponds to the general case of $\Gsym$ acting homogeneously on the internal manifold $\Mint$. Whenever such an action exists, it is always true that $\Mint \cong \Gsym/H$, where $H$ is the stabiliser of $\Mint$ under the action of $\Gsym$. Hence, in these cases $ \Mint / ( \Gsym/H) $ will be trivial and the tower of singlets will be finite. The coset matrix $L(y)$ will act as a change of basis, twisting the algebra of singlets expressed in terms of the global $\Edd$ action carried by $b\in \mathcal{C}_{\Edd}\left(H\right)$. 
Furthermore these cases will lead to finite consistent truncations, and it is in the coordinate-independent basis where we will be able to identify the scalars of the truncated supergravity as the $x$-dependent parameters appearing in the expansion of $b$ in terms of generators.

\paragraph{$\mathrm{SO}(7)_v$ symmetry (inhomogeneous case).} 
Under $\mathrm{SO}(7)_v$ we have the following branching of the fundamental representations of $\mathrm{SO}(8)$:
\begin{equation}
\mathbf{8}_v  \rightarrow \mathbf{7} + \mathbf{1}\,, \quad
\mathbf{8}_c  \rightarrow \mathbf{8}  \,, \quad
\mathbf{8}_s  \rightarrow \mathbf{8}  \,.
\end{equation}
Correspondingly, we further branch under $\mathrm{SO}(7)_v$ the fundamental $\mathbf{8}_v$ indices, $\as=(i,8)$ with $i=1\,, \dots \,,7$ transforming in the $\mathbf{7}$ of $\mathrm{SO}(7)_v$ and the $8$ index transforming as a singlet. This branching carries through for $\Gseven$ generators $R_{\as\bs}$, $S_{\as \bs}$, $t^\pm_{\as\bs\cs\ds}$ of \eqref{Rstpm}.

The singlets under $\mathrm{SO}(7)_v$ are then obtained by restricting $S^{\alpha\Sigma}$ to be an $\mathrm{SO}(7)_v$ invariant tensor. 
In this case, as the action of $\Gsym$ is not homogeneous, the embedding coordinate $\mu^8$ is a singlet by itself.
Hence any product of $\mu^8$ with another singlet will be a singlet as well. 
We therefore already see that we will encounter an infinite number of singlets, which we can arrange in terms of a tower in powers of $\mu^8$.

For this reason, consider first all possible singlets which are $\mu^8$-independent. The only $\mathrm{SO}(7)_v$ invariant tensor is $\delta_{ij}$ and so there are limited possibilities to construct singlets.
The only $\mu^8$-independent ones are: 
\begin{equation}
S_{88}\,,\quad \mu^i R_{i8} \,,\quad \mu^i S_{i8}\,, \quad \mu^i \mu^j S_{ij} \,,
\label{7singletgens}
\end{equation}
again terminating at level 2 of the KK tower.
These four singlets can be seen as the generators of the full KK tower of singlets, which is obtained by taking products of \eqref{7singletgens} with arbitrary powers of $\mu^8$. 
It's illustrative to write down the first few levels: 
\be
\begin{array}{ccccc}
\text{Level 0:} & S_{88} & & & \\
\text{Level 1:} & \mu^8 S_{88} & \mu^i R_{i 8} & \mu^i S_{i 8}  & \\
\text{Level 2:} & (\mu^8)^2 S_{88} & \mu^8  \mu^i R_{i 8} & \mu^8 \mu^i S_{i 8}  & \mu^i \mu^j S_{ij} \\
\text{Level 3:} & (\mu^8)^3 S_{88} & (\mu^8)^2\mu^i R_{i 8} & (\mu^8)^2\mu^i S_{i 8}  & \mu^8  \mu^i \mu^j S_{ij} \\
\vdots & & & & 
\end{array} 
\ee
Now, it is easy to check that the generators \eqref{7singletgens} close into an algebra with non-constant structure coefficients. 
We expect that this algebra should be linked to the imposition of a generalised $\mathrm{SO}(6)_v$ structure.
By considering coordinate-dependent combinations of the generators, we can construct the expected commutant $\mathcal{C}_{\Gseven}\left(\mathrm{SO}(6)_v\right)=\mathrm{SO}(1,1) \times \mathrm{SL}(2)$.
This is achieved using the following combinations:
\begin{equation}
H_0 = 2 \left( S_{88} + \tfrac{1}{1-(\mu^8)^2} \mu^i \mu^j S_{ij} \right)\,,
\label{SO7vSymmSO(1,1)factor}
\end{equation}
\begin{equation}
H_1 =  S_{88} - \tfrac{1}{1-(\mu^8)^2} \mu^i \mu^j S_{ij}  \,, \quad
 E_1=\sqrt{\tfrac{2}{1-(\mu^8)^2}}\, \mu^i\left( S_{i8} -  R_{i8} \right) \,, \quad  F_1= - \sqrt{\tfrac{2}{1-(\mu^8)^2}} \,  \mu^i \left( S_{i8} + R_{i8} \right)\,,
\label{SO7vSymmSL2factor}
\end{equation}
where \eqref{SO7vSymmSO(1,1)factor} generates the $\mathrm{SO}(1,1)$ factor and \eqref{SO7vSymmSL2factor} generates the $\mathrm{SL}(2)$ one. Their commutation relations are the same as the corresponding ones for the generators appearing in appendix \ref{AppendixSO7/SO6Coset}. 

These four generators can be used as a different basis to serve as the generators of the KK tower.  Note that their $\mu^8$-dependent coefficients can be Taylor expanded, so that they involve infinite sums of massive KK modes.  As there is a unique coordinate-independent realisation of $\mathrm{SO}(1,1) \times \mathrm{SL}(2)$ as the commutant of $\mathrm{SO}(6)_v$ within $E_{7(7)}$ (see appendix \ref{AppendixSO7/SO6Coset}), there must exist a coordinate-dependent matrix $L(\mu)$ such that both basis are related. 

Thus, suppose that $\tilde{b}(y)\equiv \tilde{b}(\mu(y))$ is an arbitrary linear combination (with constant coefficients) of the generators in \eqref{SO7vSymmSO(1,1)factor} and \eqref{SO7vSymmSL2factor}.
We must have $\tilde{b}(y) = L(y) b L(y)^{-1} $, where $b$ is coordinate independent.
The most general singlet then takes into account the $\mu^8$ dependence and can be expanded in the form
\be
S = L(y) b( \mu^8) L(y)^{-1} 
\,,\quad
b (\mu^8) = \sum_{n,\aadj } \phi_{n , \aadj }(x) (\mu^8)^n b_{\aadj }
\,,	
\ee
where the index $\aadj$ runs on the Lie algebra of $\mathcal{C}_{\Gseven}\left(\mathrm{SO}(6)_v\right)$. Now, in the cases where $\Gsym$ acts homogeneously, we could view $L(y)$ as a rotation of the algebra of singlets to a coordinate-independent basis. 
In the present inhomogeneous cases, $L(y)$ is used to transform to the generators $b(\mu^8)$ which form a Lie algebra with non-constant structure coefficients.
Only for the level 0 generators, which are the $b_\aadj$ above, do we get a standard Lie algebra with constant coefficients.

What are we seeing here can be recast in the language of Kac-Moody algebras. 
One way to construct such an algebra (we only consider vanishing central extension) is to start with some Lie algebra with generators $b_\aadj$ and structure constants $f_{\aadj \badj}{}^{\cadj}$.
Then introduce some variable $u$ (which for the usual loop algebra construction we take to be $u= e^{i\theta}$ with $\theta \in S^1$) and consider the algebra generated by all tensors products $u^n \otimes b_{\aadj}$.
This has the form
\begin{equation}
[ u^n b_{\aadj}, u^m b_{\badj} ] = u^{n+m} f_{\aadj \badj}{}^{\cadj} b_{\cadj} \,.
\end{equation}
We can refer to the generators $u^n \otimes b_{\aadj}$ as the level $n$ generators, and the original Lie algebra occurs at level 0.

Our singlet construction gives rise in this case to such an algebra with $u = \mu^8$. The generators $b_a$ correspond to those obtained by factoring the generators of the singlet tower (i.e. \eqref{SO7vSymmSO(1,1)factor} and \eqref{SO7vSymmSL2factor}) using the matrix $L(y)$ to obtain a coordinate independent basis.
Moreover, considering the sphere parametrisation \eqref{SO7vEmbeddingCoordinates} we can see that $\mu^8=\cos(\theta_7)$ can be expanded in terms of exponentials of the kind $e^{i n \theta_7}$, so that the connection with the loop algebra definition through $S^1$ harmonics is direct. The KK tower expanded in powers of $\mu^8$ can be re-summed just by requiring that $b(\theta_7)$ in \eqref{SingletSolution} is an arbitrary function in $C^\infty \left( S^1 \right) \otimes \mathcal{C}_{\Gseven}\left(\mathrm{SO}(6)_v\right)$. Note that in this case $\Mint / (\Gsym/H)$ is one dimensional and compact, so it can only be $S^1$. It is worth remarking that for inhomogeneous cases, $L(y)$ contains a sum over an infinite number of KK modes, as it can be seen for the example worked out here if \eqref{SO7vL(y)8v} is expanded in terms of $\mathrm{SO}(8)$ harmonics.

\paragraph{Generalised Kac-Moody algebras.} The two examples discussed above can be summarised in a more general framework. As shown in the previous section, in general singlets are characterised by a coset $L(y)$ and $\mathcal{C}_{E_{d(d)}}(H)$-valued functions on $\Mint / (\Gsym/H)$ we denote by $b(y)$ (see \eqref{SingletSolution}). These $\mathcal{C}_{E_{d(d)}}(H)$-valued functions $b(y)$ can be expanded in terms of harmonics of $\Mint / (\Gsym/H)$, say $\{ g_{\mathcal{I}}(y) \}$, and generators of $\mathcal{C}_{E_{d(d)}}(H)$, say $\{b_a\}$. Then the commutation relations give a \emph{generalised Kac-Moody algebra} of the type studied in \cite{Dolan:1983aa,Campoamor-Stursberg:2021iuq}, and can be written as:
\begin{equation}
[g_{\mathcal{I}}(y) b_a\, , \, g_{\mathcal{J}}(y) b_b]= f_{\mathcal{I}\mathcal{J}}{}^\mathcal{K} f_{ab}{}^c\, g_{\mathcal{K}}(y) b_c\,,
\label{KacMoodyAlgebraCommutationRelationships}
\end{equation}
where $f_{\mathcal{I}\mathcal{J}}{}^\mathcal{K} g_{\mathcal{K}}(y)$ accounts for the harmonic expansion of the product of two harmonics. From the level zero harmonic we recover the Lie algebra used to construct the Kac-Moody algebra. The role of $L(y)$ is then to make this structure manifest at the level of the infinitesimal singlets which one has to exponentiate to retain all the $\Gsym$ invariant KK modes. For homogeneous actions on $\Mint$ the manifold $\Mint / (\Gsym/H)$ is trivial and one recovers a honest Lie algebra, since the only functions that can be defined over the set $\Mint / (\Gsym/H)$ containing just one element are constant functions.

\section{Infinite and finite truncations on the seven-sphere} \label{sec:examples}

\subsection{Generalised parallelisations on $S^7$} 
\label{GenParallelizationsSevenSphere}

11-dimensional supergravity admits a consistent truncation on $S^7$ to the $\mathrm{SO}(8)$ gauged maximal supergravity in four dimensions.
This can be described as a generalised parallelisation in $\Gseven$ exceptional geometry.
Let's first introduce some geometric notions associated with the round $S^7$.
The metric $\gr_{ij}$ in some coordinates $y^i$ is defined by
\be
\dd s^2 = \gr_{ij} \dd y^i \dd y^j = R^2 \delta_{\as \bs} \dd \mu^\as \dd \mu^\bs\,,
\ee
where $\mu^\as$, $\as = 1,\dots,8$, are embedding coordinates in $\mathbb{R}^8$, defining the unit sphere via $\delta_{\as \bs} \mu^\as \mu^\bs=1$.
There is a flux of the seven-form field strength with $F_{(7)} = \tfrac{6}{R} \mathrm{vol}_{\gr}$, where the volume form is 
\be
\mathrm{vol}_{\gr} = \tfrac{R^7}{7!} \epsilon_{\as_1 \dots \as_{8}} \mu^{\as_1} \dd \mu^{\as_2} \wedge \dots\wedge \dd \mu^{\as_{8}} \,.
\ee
The sphere admits conformal Killing vectors $k_\as$ such that $L_{k_\as} \gr_{ij} = - 2 \mu_\as \gr_{ij}$ and $L_{k_\as} \mu_{\bs} = \delta_{\as \bs} - \mu_{\as} \mu_{\bs}$. In terms of these, the Killing vectors and the $\mathrm{SO}(8)$ Lie algebra they generate are: 
\be
v_{\as \bs} = R^{-1} ( \mu_\as k_\bs - \mu_\bs k_\as ) \,,
\quad
L_{v_{\as \bs}} v_{\cs \ds} =4  R^{-1} \delta_{\as][\cs} v_{\ds][\bs}\,.
\ee
In $\Gseven$ exceptional geometry, the generalised tangent bundle is $\mathcal{E} \approx TM \oplus \Lambda^2 T^*M \oplus \Lambda^5 T^*M \oplus (T^*M \otimes \Lambda^7 T^*M)$, so that a generalised vector $V^M$ sits in the fundamental representation $R_1 = \mathbf{56}$ of $\Gseven$.
This decomposes under $\mathrm{SL}(8)$ into the $\mathbf{28} \oplus \mathbf{\overline{28}}$, thus $V^M = (V^{\fM\fN}, V_{\fM \fN})$, where $\fM=1,\dots,8$ and the pairs of indices $\fM\fN$ are antisymmetric.

The generalised parallelisation on $S^7$ was worked out in \cite{Lee:2014mla}.
The generalised metric is decomposed as 
\be
\gM_{MN} = \Ur_M{}^A \delta_{AB} \Ur_N{}^B  \,,
\ee
in terms of a globally defined generalised vielbein which decomposes in terms of $\mathrm{SL}(8)$ representations as
\be
\Ur_M{}^A = \begin{pmatrix} 
 2  \Ur_{[\fM}{}^{\as}  \Ur_{\fN]}{}^{\bs}& 0 \\
 0 &2  \Ur^{\fM}{}_{[\as}  \Ur^{\fN}{}_{\bs]}
 \end{pmatrix}\,,
\label{E7U}
\ee
where, letting $\fM= (i,8)$, we have $\mathrm{SL}(8)$ matrices 
\begin{align} 
 \Ur_\fM{}^\fA 
& = 
%{g}^{\tfrac14 \tfrac{d-3}{d+1}} 
\gr^{\tfrac{1}{8}}
\begin{pmatrix} 
R \gr^{-1/4} \partial_i \mu^\fA \\
- \gr^{1/4} \mu^\fA
+ R \gr^{-1/4} C^i \partial_i \mu^\fA
\end{pmatrix} 
\,,\quad
\Ur^{\fM}{}_{\fA} 
 = %{g}^{-\tfrac14 \tfrac{d-3}{d+1}}
\gr^{-\tfrac{1}{8}} 
\begin{pmatrix}
R^{-1} \gr^{1/4} k^i{}_{\fA} + \gr^{-1/4} C^i \mu_{\fA}\\
- \gr^{-1/4} \mu_{\fA} 
\end{pmatrix} \,,
\end{align}
which are inverse of each other, i.e. $\Ur_\fM{}^\fA \Ur^\fN{}_\fA = \delta_\fM^\fN$.
Here we use a six-form gauge potential $C_{i_1 \dots i_6}$ for the seven-form field strength, and rewrite this as a vector density $C^i = \tfrac{1}{6!} \epsilon^{ii_1 \dots i_6} C_{i_1 \dots i_6}$, such that $\partial_i C^i = \tfrac{6}{R} \gr^{1/2}$.
In addition, the scalar density $\Delta$ is given by $\Delta = \gr^{1/4}$.
Letting $\Urv_A = \Delta \Ur_A$, the 
non-trivial part of the algebra is:
\be
\mathcal{L}_{\Urv_{\as \bs}} \Urv_{\cs \ds}  = R^{-1} 4 \delta_{\as] [ \cs} \Urv_{\ds][\bs}  \,,\quad
\mathcal{L}_{\Urv_{\as \bs}} \Urv^{\cs \ds}  = R^{-1} 4 \delta_{[\as}^{[\cs} \delta_{\bs] \es} \Urv^{\ds] \es} \,,
\ee
realising the $\mathrm{SO}(8)$ gauging with
\be
X_{\as \bs, \cs \ds}{}^{\es \fF}=
- X_{\as\bs}{}^{\es\fF}{}_{\cs \ds}  = - 8 R^{-1} \delta_{[\as}^{[\es} \delta_{\bs][\cs} \delta_{\ds]}^{\fF]} \,.
\ee	
We see that $\Mint = S^7$ is an example of a manifold admitting a generalised parallelisation leading to a consistent truncation, where $\Ggauging = \mathrm{SO}(8)$ corresponds to the isometries of the background.

We can find generalised parallelisations preserving some symmetry $\Gsym \subseteq \mathrm{SO}(8) \subset \mathrm{SL}(8) \subset \Gseven$ following the general logic of section \ref{logic}. 
For a Lie algebra generator $a \in \Gsym$, its embedding is explicitly given by:
\be
a = a^A X_A{}^\alpha t_\alpha 
= \tfrac12 a^{\as \bs} X_{\as \bs}{}^\cs{}_\ds t_{\cs}{}^\ds 
= R^{-1} a^{\as \bs} t_{[\as \bs]} 
\ee
where we used the explicit form of the $\mathrm{SO}(8)$ embedding tensor $X_{\as \bs}{}^{\cs}{}_{\ds} = 2 R^{-1} \delta^{\cs}_{[\as} \delta_{\bs]\ds}$ as well as the $\mathrm{SL}(8)$ decomposition of the adjoint, $t_\alpha = ( t_\as{}^\bs, t_{\as \bs \cs \ds})$.
The generators $t_{\as}{}^{\bs}$ give the $\mathbf{63}$ of $\mathrm{SL}(8)$.
Lowering the second index with $\delta_{\as \bs}$ we have $R_{\as \bs} = 2 t_{[\as\bs]}$ giving the $\mathbf{28}$ of $\mathrm{SO}(8)$ and $S_{\as \bs} = 2 t_{(\as\bs)}$ giving the $\mathbf{35}$, as in \eqref{Rstpm}.
We then seek to construct the singlets obeying the equation \eqref{SingletEquation}, which can be written as
\be
\tfrac12 
a{}^{\as \bs} L_{v_{\as \bs}} S = \tfrac12 a^{\as \bs} [ R_{\as \bs}, S ]
\ee
taking the Killing vectors to correspond to those of the unit sphere to eliminate an	 overall factor of $R^{-1}$.
The singlets solving this equation will have the form of \eqref{SingletSolution} and the corresponding generalised parallelisations compatible with $\Gsym$ can be defined by \eqref{defMU} (note $\Delta$ is unchanged).

\subsection{Example: finite truncation with $G_{\text{sym}}=\mathrm{SU}(4)$} \label{sec:ExampleSU4}

For our first example, we choose $G_{\text{sym}}=\mathrm{SU}(4)_c$.
We already discussed some aspects of this case in section \ref{SingletTowers}.
There we noted that $\mathrm{SU}(4)_c$ acts homogeneously on $S^7$, so that $S^7\cong \mathrm{SU}(4)/\mathrm{SU}(3)$ as an homogeneous space, the stabiliser of each point of $S^7$ being isomorphic to $\mathrm{SU}(3)$.
Hence the group $H$, that becomes the generalised structure group, is $H=\mathrm{SU}(3)$. 
The generators $b$ then lie in $\mathcal{C}_{E_{7(7)}}\left( \mathrm{SU}(3) \right) = \mathrm{SL}(2) \times \mathrm{SU}(2,1)$ and are independent of the sphere coordinates. 
Accordingly, all fields that will appear in the consistent truncation depend only on the four-dimensional coordinates $x$.

Applying our construction in this case leads to a consistent truncation first worked out in \cite{Gauntlett:2009zw} as a reduction of M-theory on a 7-dimensional Sasaki-Einstein manifold, $SE_7$. This consistent truncation led to a four-dimensional $\mathcal{N}=2$ supergravity with five scalars, two vector fields and a two-form, which can be dualised into another scalar.  

Our approach will give an explicit embedding of this $\mathcal{N}=2$ consistent truncation into the generalised parallelisation on the seven-sphere. 
Note that by starting with the duality-covariant $\Gseven$ ExFT we will in fact retain six scalars, four vectors and two two-forms: the equations of motion and Bianchi identities of ExFT will encode the possible electromagnetic duality relations between them (for instance, as usual two of the vectors can be chosen to be electric and the other two magnetic). 

The original consistent truncation of \cite{Gauntlett:2009zw} was extended in \cite{Gauntlett:2009bh} by introducing a sign parameter $\epsilon=\pm 1$ appearing in the ansatz for the four-form flux, and corresponding to skew-whiffing (see \cite{Duff:1986hr}) of the underlying AdS${}_4 \times SE_7$ solutions.
More precisely, we  explain in this section how requiring $\Gsym=\mathrm{SU}(4)_c$ leads to the truncation with $\epsilon=-1$ (with action given by equation (2.6) of \cite{Gauntlett:2009bh}).
The truncation with $\epsilon=+1$ corresponds to the case $G_{\text{sym}}=\mathrm{SU}(4)_s$, which can be worked out analogously.

The ingredients we use for this embedding, and the field content involved, are summarised in table \ref{su4stuff}.
Many further explicit details can be found in appendix \ref{appSU4}.

\begin{table}[h]
\centering
\renewcommand{\arraystretch}{1.2}

\begin{tabular}{|c|c|c|}
\hline
\multicolumn{3}{|c|}{$\Gsym = \mathrm{SU}(4)_c$ \hspace{1em} $H= \mathrm{SU}(3)$} \\ \hline
Parallelisation & $U_A = \mathring{U}_B (L^{-1})_A{}^B$ &   \\ & $L \in \mathrm{SU}(4)_c/ \mathrm{SU}(3)$ & \eqref{SU4cL(y)8v} and \eqref{SU4cL(y)56from8v}\\\hline
Scalars & $\mathcal{C}_{\Gseven}(\mathrm{SU}(3)) = \mathrm{SL}(2) \times \mathrm{SU}(2,1)$ & \\
 & $\mathcal{V}(x)\in \tfrac{ \mathrm{SL}(2) \times \mathrm{SU}(2,1)}{\mathrm{SO}(2) \times\mathrm{SU}(2) \times \mathrm{U}(1)}$ & \eqref{SU4cCosetParametrization}\\
 & $\phi^I = ( U,V,h,a, \chi_1,\chi_2 )$ & \\ \hline
Vectors & 4 invariant $K_{\ivec }$  &   \eqref{SU3InvariantVectorsExplicit}\\ 
 & $A_\mu = K_{\ivec } A_\mu{}^{\ivec }$ & \\ \hline
2-forms & 11 invariant $J^{\aadj }$ & Generators of $\mathrm{SL}(2) \times \mathrm{SU}(2,1)$ \\
 & $B_{\mu\nu} = J^{\aadj } B_{\mu\nu \aadj }$ & Only 2 appear
\\ \hline
\end{tabular}

\caption{Ingredients for the $\mathrm{SU}(4)_c$ symmetric deformation of the $S^7$ generalised parallelisation. This leads to a finite-mode consistent truncation.} 
\label{su4stuff} 
\end{table}

We start by constructing the $G_{\text{sym}}=\mathrm{SU}(4)_c$ singlets. The group $\mathrm{SU}(4)_c$ corresponds to the $\mathrm{SU}(4) \subset \mathrm{SO}(8)$ which branches the three fundamental representations of $\mathrm{SO}(8)$ according to \eqref{SO8branchSU4}.
The generators on the $\mathbf{8}_v$ representation are given in equation \eqref{SU4cGenerators8v}. 

In accordance with the KK analysis of section \ref{SingletTowers}, we find that all the singlets are of the form \eqref{SingletSolutionExponential} with $L(y)$ parametrising the coset $\mathrm{SU}(4)/\mathrm{SU}(3)\cong S^7 $ and $\mathcal{V}(x)$ being generated by $\mathcal{C}_{E_{7(7)}}\left( \mathrm{SU}(3) \right) = \mathrm{SL}(2) \times \mathrm{SU}(2,1)$.
The simplicity of the computations heavily relies on the form of $L(y)$ which, for this particular case, takes care of all the internal coordinate dependencies. We provide a simple parametrisation for $L(y)$ in terms of  `Euler angles' in equations \eqref{SU4cL(y)8v} and \eqref{SU4cL(y)56from8v}, the latter being its $56$ dimensional representation constructed from the former. 

In appendix \ref{appSU4}, we provide the explicit generators of the generalised $\mathrm{SU}(3)$ structure as well as its commutant.
To perform the consistent truncation, we also need to introduce the $\mathrm{SU}(3)$ invariant generalised vectors $K_{\ivec }$, $\ivec =1,2,3,4$, defined in \eqref{SU3InvariantVectorsExplicit}.

The generalised parallelisation defined by our choice \eqref{SU4cL(y)56from8v} of $L(y)$ will have non-constant generalised fluxes defined by \eqref{IntrinsicTorsionComponentsForNonMaximalTruncations}. 
Crucially, we find by explicit computation that though the full intrinsic torsion $X_{AB}{}^C$ is not constant, the components $X_{\ivec B}{}^C$ of the intrinsic torsion which will be singled out by the $\mathrm{SU}(3)$ invariant gauge fields are constant and lie in the commutant of $\mathrm{SU}(3)$.  
Furthermore, we find that $\theta_{\ivec }=0$ even though $\theta_A$ is not zero, so the truncated theory does not gauge the trombone symmetry.
However, the non-vanishing components of the full trombone do play a role in the effective potential.
The upshot is that we indeed meet the general conditions to have a finite consistent truncation, with generalised $\mathrm{SU}(3)$ structure.

Now let us discuss how we see the field content of this consistent truncation.
The coset $\mathcal{V}(x)$ will parametrise six scalar fields. Its explicit form is given in \eqref{SU4cCosetParametrization}, where we denote the scalar fields as
\begin{equation}
\phi^\mathcal{I}=
\left(
U\, , \, V \, ,\,
h\, ,\,
a\, ,\,
%\zeta
\chi = \chi_1 + i \chi_2
\right)\,.
\end{equation}
The vector fields and two-forms are expanded following the ansatz \eqref{ABexpand}. 
We have four vector fields $A_\mu{}^{\ivec }$, however the field strengths of these four vectors will be related by the four-dimensional self-duality constraint, and below we will pick two to be regarded as the electric vector fields that appear in the dynamics.
Computing explicitly, we find that while the gaugings $X_{\ivec  A}{}^B$ are non-zero in general, the combination $X_{\ivec \jvec }{}^{\kvec }$ from \eqref{GaugeFieldsGeneralizedDerivativeClosure} vanishes such that the vector fields generate an abelian gauge algebra.
Note that even though the lower dimensional structure coefficients $X$ are vanishing, we will find non-vanishing covariant derivatives in the scalars, so the gauging cannot be fully characterised in terms of $X_{\ivec \jvec }{}^{\kvec }$. 
 
The field strengths of the one-forms include contributions from the two-forms, which in $\Gseven$ ExFT are Lie algebra valued.
The adjoint valued $\mathrm{SU}(3)$ invariants $J^{\aadj }$ correspond to the generators of $\mathcal{C}_{E_{7(7)}}( \mathrm{SU}(3) )$, given in \eqref{SU4cCommutantGeneratorsSL2} and \eqref{SU4cCommutantGeneratorsSU(2,1)}. 
From \eqref{Fpartway}, these appear in the field strength projected by a particular combination of the generalised fluxes, $( X^{A \alpha}  - 24 t^{\alpha AB } \theta_B ) J^{\aadj }{}_{\alpha}$, which here as required turns out to be constant and $\mathrm{SU}(3)$ invariant. 
\setcounter{MaxMatrixCols}{11}
Explicitly we find that the ensuing field strength \eqref{FieldStrengthsTruncation} involves the matrix \begin{equation}
\small
\Omega^{\ivec \jvec }K_{\jvec }{}^A( X_{AB}{}^{C}  - 24  \mathbb{P}^C{}_B{}^D{}_A \theta_D )J^{\aadj }{}_C{}^B= 
\begin{pmatrix}
0 & 0 & 0 & 0 & 0 & 0 & 0 & 0 & 0 & 0 & 0 \\
0 & 0 & 0 & 0 & 0 & 0 & 0 & 0 & 0 & 0 & 0 \\
0 & 0 & 0 & 0 & 0 & 0 & 0 & 0 & 0 & \frac{24 \sqrt{6}}{R} & 0 \\
0 & 0 & 0 & 0 & -\frac{96}{R} & 0 & 0 & 0 & 0 & -\frac{72 \sqrt{2}}{R} & 0
\end{pmatrix}
\,.
\label{ThetaBThing}
\end{equation}
This projects out all except two of the two-forms.
After the redefinitions $B_{(10)} \rightarrow  \frac{\sqrt{2}}{ 24} B_{(1)} \,, B_{(5)} \rightarrow \frac{1}{96} B_{(2)} $ of the two-forms, the field strengths are:
\begin{equation}
F^{\ivec }=dA^{\ivec } + B^{\ivec }
= \begin{pmatrix}
d A^{(1)} \\
d A^{(2)}\\ 
d A^{(3)} + \frac{2 \sqrt{3} B_{(1)}}{R}\\
d A^{(4)}-\frac{ B_{(2)}}{R}-\frac{6  B_{(1)} }{R} \,.
\end{pmatrix} 
\label{SU4Fs}
\end{equation}
We have to impose the self-duality condition, \eqref{SelfDuality_CT_Recast}.
In components, this requires making a choice of which two vectors are electric and which are magnetic.
To guide us in this choice, we now look in detail at the equations of motion and how they compare to those of \cite{Gauntlett:2009zw, Gauntlett:2009bh}.

\vspace{1em}	
\noindent\emph{Einstein equation.} 
Let us first discuss the different terms appearing in the Einstein equation \eqref{ExcEin_CT}. 
We can immediately note the Ricci tensor and scalar are their genuine four-dimensional selves, as the trombone $\theta_{\ivec }$ is zero.
The effective potential \eqref{Veff_CT} is determined in terms of the full non-constant embedding tensor and trombone of the deformed generalised parallelisation. 
This is a function which geometrically is a singlet of $\mathrm{SU}(4)_c$ and so obeys $\tfrac12 
a{}^{\as \bs} L_{v_{\as \bs}} V_{\text{eff}}=0$ for all the $\mathrm{SU}(4)_c$ generators. 
This requires that it be constant. Intuitively, this can be shown from the fact that the action of $\mathrm{SU}(4)_c$ on $S^7$ is homogeneous, so there are no fixed points, nor combinations of embedding coordinates which are annihilated by $\tfrac12 
a{}^{\as \bs} L_{v_{\as \bs}}$. 
Explicitly, evaluating the expression of the effective potential \eqref{Veff_CT} in terms of the deformed generalised parallelisation, we find
\begin{equation}
\begin{split}
V_{\text{eff}}&= \frac{1}{R^2} \Big(24 h^2 e^{-14 U - V}+24 |\chi|^2 e^{-12 U-3 V}+6 e^{-10 U + V}-48 e^{-8 U -V}\\
& \quad \quad \quad \quad +18 \left(h^2+|\chi|^2-1\right)^2 e^{-18U -3 V} \Big) \,.
\end{split}
\label{VeffSU4}
\end{equation}
For $R=1$, this matches the potential found in \cite{Gauntlett:2009zw, Gauntlett:2009bh}.

We stress that each of the individual terms appearing in the effective potential \eqref{Veff_CT} depends on the internal coordinates, even those involving the trombone $\theta_A$, and it is only the full effective potential what turns out to be constant. This highlights the fact that even though there is no gauging of the trombone symmetry in the $\mathcal{N}=2$ supergravity arising from the truncation, the $D=11$ trombone-like terms $\theta_A$ are essential for the potential to be constant. 

Next we can consider the terms in the Einstein equation \eqref{ExcEin_CT} involving the scalars and the gauge fields.
After redefining $A^{(1)} \rightarrow - \sqrt{3} A^{(1)} $, we find:
\begin{equation}
\tfrac{1}{4 \alpha} {D}_\mu M_{AB} {D}_\nu M^{AB}  = \mathcal{G}_{\mathcal{I} \mathcal{J}}  D_\mu \phi^\mathcal{I}  D_\nu \phi^\mathcal{J}
\end{equation}
where the metric on the scalar field space is
\begin{equation}
\small
 \mathcal{G}_{\mathcal{I} \mathcal{J}}=
 \left(
\begin{array}{cccccc}
 -24 & -3 & 0 & 0 & 0 & 0 \\
 -3 & -\frac{3}{2} & 0 & 0 & 0 & 0 \\
 0 & 0 & -\frac{3}{2}  e^{-2 (2U+V)} & 0 & 0 & 0 \\
 0 & 0 & 0 & -\frac{1}{2} e^{-12U} & \frac{3}{4} \chi_2 e^{-12U} & -\frac{3}{4}  \chi_1  e^{-12U} \\
 0 & 0 & 0 & \frac{3}{4} \chi_2 e^{-12U} & -\frac{3}{8}  e^{-12U} \left(3 \chi_2^2+4 e^{6U}\right) & \frac{9}{8} \chi_1  \chi_2 e^{-12U} \\
 0 & 0 & 0 & -\frac{3}{4}  \chi_1  e^{-12U} & \frac{9}{8} \chi_1  \chi_2 e^{-12U} & -\frac{3}{8}  e^{-12U} \left(3 \chi_1 ^2+4 e^{6U}\right) \\
\end{array}
\right)\,,
\end{equation}
and the covariant derivatives are defined as:
\begin{equation}
D \phi^{\mathcal{I}}=
\left( 
dU\,, \,\, dV \,, \,\, dh \,, \,\, da +\tfrac{6}{R} \left( A^{(1)}+A^{(2)} \right) \,, \,\, d\chi_1 + \tfrac{4  }{R} \chi_2 A^{(2)} \,, \,\, d\chi_2-\tfrac{4   }{R}\chi_1 A^{(2)}
\right)\,.
\label{ScalarCovDerExplicit}
\end{equation}
This expression characterises the kinetic terms of the scalars, as well as their coupling to the  $\mathrm{SU}(3)$ invariant vectors. The term $\tfrac{1}{4 \alpha} {D}_\mu M_{AB} {D}^ \mu M^{AB}= \mathcal{G}_{\mathcal{I} \mathcal{J}}  D_\mu \phi^\mathcal{I} D^\mu \phi^\mathcal{J}$ exactly matches the kinetic term of the scalars and their coupling to the vector fields appearing in the lagrangian (2.6) of \cite{Gauntlett:2009bh} for the choice $\epsilon = -1$ under the identifications:
\begin{equation}
\quad A^{(1)} =\tilde{B}_1^{\text{there}} \, , \quad A^{(2)} = A_1^{\text{there}} \, ,
\end{equation}
in terms of the one forms $\tilde B_1$ and $A_1$ of \cite{Gauntlett:2009zw, Gauntlett:2009bh}.
Consequently, as the contributions involving scalar fields in the Einstein equation \eqref{ExcEin_CT} coincide with what one obtains from the variation of the non-linear sigma model specified by $\mathcal{G}_{\mathcal{I} \mathcal{J}}  D_\mu \phi^\mathcal{I} D^\mu \phi^\mathcal{J}$, we find that they match what one would obtain from the variation of the Lagrangian (2.6) of \cite{Gauntlett:2009bh}. Note that these identifications have fixed $A^{(1)}$ and $A^{(2)}$ as the vectors that must be chosen to be the electric ones in order to obtain the truncation in the same form as in \cite{Gauntlett:2009bh}. 

Referring to \eqref{SU4Fs}, we see that $F^{(1)}$ and $F^{(2)}$ are abelian and have no contribution from the two-forms, which implies the identifications:
\begin{equation}
\quad F^{(1)} =\tilde{H}_2^{\text{there}} \, , F^{(2)} = F_2^{\text{there}} \, ,
\end{equation}
in terms of the field strengths defined in \cite{Gauntlett:2009zw,Gauntlett:2009bh}. 

Consider now the self-duality condition  \eqref{SelfDuality_CT_Recast}. 
This involves the following projection of the scalar matrix $ M_{\ivec \jvec } = K_{\ivec } {}^A M_{AB}  K_{\jvec } {}^B $.
Using the parametrisation \eqref{SU4cCosetParametrization} and the definition of the invariant vectors \eqref{SU3InvariantVectorsExplicit}, we obtain
\begin{equation}\scriptsize
\begin{array}{l}
e^{6 U + 3V}M_{\ivec \jvec }= 
\left(
\begin{array}{cccc}
 3 h^2+e^{4 U+2 V} & \sqrt{3} h^2 \left(h^2+e^{4 U+2 V}\right) & h \left(3 h^2+2 e^{4 U+2 V}\right) & \sqrt{3} h \\
 \sqrt{3} h^2 \left(h^2+e^{4 U+2 V}\right) & \left(h^2+e^{4 U+2 V}\right)^3 & \sqrt{3} h \left(h^2+e^{4 U+2 V}\right)^2 & h^3 \\
 h \left(3 h^2+2 e^{4 U+2 V}\right) & \sqrt{3} h \left(h^2+e^{4 U+2 V}\right)^2 & 3 h^4 +4 h^2 e^{4 U+2V}+e^{8 U+4V} & \sqrt{3} h^2 \\
 \sqrt{3} h & h^3 & \sqrt{3} h^2 & 1 \\
\end{array}
\right)
\end{array}\,.
\end{equation}
This is simply an $\mathrm{SL}(2)/\mathrm{SO}(2)$ coset scalar matrix in the four-dimensional representation. 
In terms of the $\mathcal{N}=2$ consistent truncation, this matrix codifies the coupling between the vectors and the scalars in the vector multiplet.

We now use the self-duality condition to express $F^{(3)}$ and $F^{(4)}$ in terms of $F^{(1)}$ and $F^{(2)}$. 
The explicit solution reads:
\begin{equation}
\small
\begin{split}
F^{(3)} &= \frac{1}{4 h^2+e^{4 U+2 V}} \left( 2\sqrt{3} h F^{(1)} - \sqrt{3} h \left( e^{4 U+2 V} +2 h^ 2 \right)  F^{(2)} - \sqrt{3} 
       e^{2 U + V} *(F^{(1)} + h^2 F^{(2)})  \right)\,,\\
F^{(4)} &=    \frac{1}{4 h^2+e^{4 U+2 V}}
       \Bigg( 3 h \left( e^{4 U+2 V} +2 h^ 2 \right) F^{(1)} +3 h^ 2 
         e^{2 U + V} *F^{(1)} \\
         & \quad \quad \quad \quad \quad \quad \quad \quad \quad \quad + \left( e^{4 U+2 V} + h^ 2 \right) \left( 2h^ 2 F^{(2)} + e^{2 U+ V} \left( e^{4 U+2 V} + 3h^ 2 \right) * F^{(2)}  \right) \Bigg) \,.
\end{split}
\label{SU4SelfDualitySolution}
\end{equation}
A tedious but straightforward computation shows that upon the imposition of the solution of the self-duality equation \eqref{SU4SelfDualitySolution}, the contribution of the vector fields to the Einstein equation reproduces the corresponding contributions of the vector fields to Einstein's equation directly derived from the variation of (2.6) of \cite{Gauntlett:2009bh}. All the contributions worked out above show that upon the truncation ansatz of the beginning of the section \eqref{ExcEin_CT} reproduces Einstein's equation of motion derived from the variation of (2.6) of \cite{Gauntlett:2009bh}.

\vspace{1em}
\noindent \emph{One-form equation.}  We now turn to the one-form equation of motion \eqref{OneFormEquationForHstructReduced}, which for $\Gseven$ and with $X_{\ivec \jvec }{}^{\kvec }=0=\theta_{\ivec }$ reads
\begin{equation}
0  = |\bar g|^{-1/2} \partial_\nu ( |\bar g|^{1/2} M_{\ivec \jvec } F^{\nu\mu \jvec } ) 
+ \tfrac{1}{12} M^{CD} D^\mu M_{BD} \left( 
X_{\ivec C}{}^B - 24 K_{\ivec }{}^B \theta_C 
\right)\,,
\label{SU4OneFormFieldEquation}
\end{equation}
We evaluate the second term using the fact that $M^{CD} D^\mu M_{BD} $ is  $\mathcal{C}_{E_{7(7)}}\left( \mathrm{SU}(3) \right)$-valued: it is the same operator that appears with the two-form contributions to the field strength of the one-forms. Explicitly, one obtains:
\begin{equation}
\begin{array}{l}
\frac{1}{12}M^{CD} D_\mu M_{BD} \left( 
X_{\ivec C}{}^B - 24 K_{\ivec }{}^B \theta_C 
\right)=\\
\\
\left(
\begin{matrix}
2\sqrt{3}\frac{ e^{-12 U}}{R}\left( Da -\frac{3 i}{4} \left( {\chi}^* D{\chi} - {\chi} D{\chi}^* \right) \right)\\
6\frac{ e^{-12 U}}{R} \left( \left(|\chi|^2 - 1\right) \left( Da -\frac{3 i}{4} \left( {\chi}^* D{\chi} - {\chi} D{\chi}^* \right) \right)- i  e^{6 U}  \left( {\chi}^* D{\chi} - {\chi} D{\chi}^* \right) \right)\\
0\\
0
\end{matrix}
\right)
\end{array}\,,
\label{SecondTermEvaluatedExplicitly}
\end{equation}
 Upon the substitution \eqref{SU4SelfDualitySolution}, we find:
\begin{equation}
\begin{array}{l}
 M_{\ivec \jvec } F^{\jvec } = \\
\\
\left(
\begin{matrix}
\sqrt{3} h *F^{(2)} -\frac{\sqrt{3} }{4 h^2+e^{4 U+2 V}} \left[
e^{2 U+V} \left( F^{(1)} + h^ 2 F^{(2)} \right) + h \left( 2 *\left( F^{(1)}+h^2 F^{(2)} \right)\right)
 \right] \\
-3 h *F^{(1)}+e^{6 U+3 V} F^{(2)}-2 h^3 *F^{(2)}+ \frac{3  h^2}{4 h^2+e^{4 U+2 V}} \left[ e^{2 U+V} \left(F^{(1)}+h^2 F^{(2)} \right)+2 h *\left( F^{(1)} +  h^2 F^{(2)}  \right) \right] \\
-\sqrt{3} * F^{(1)}\\
*F^{(2)}
\end{matrix}
\right)
\end{array}\,.
\end{equation}
\begin{equation}
*d\left(  M_{\ivec \jvec } * F^{ \jvec } \right)= -  \frac{1}{\sqrt{|g|}} \partial_\nu \left( \sqrt{|g|} M_{\ivec \jvec }  F^{\nu \mu \jvec }\right) g_{\mu \rho} dx^\rho\,.
\end{equation}
It follows immediately from this that the final two components of \eqref{SU4OneFormFieldEquation} are just the Bianchi identities of the electric vectors: $dF^{(1)}=dF^{(2)}=0$. 
It is then straightforward, albeit tedious, to show that the remaining two equations of motion for the vector fields coincide with those of \cite{Gauntlett:2009bh}.

At this point we can use the one-form equations of motion together with the self-duality condition to find duality relations between the two-forms and scalar fields.
From the definition of the field strengths \eqref{SU4Fs}, since the field strengths are abelian, we can compute their Bianchi identities just by taking an exterior derivative  $dF^{\ivec }=d B^{\ivec }$, where $B^{\ivec } = (0,0,2 \sqrt{3} B_{(1)}/ R, - B_{(2)} - 6  B_{(1)}/R)$. Consequently, applying $*d$ to the self-duality condition \eqref{SelfDuality_CT_Recast} we obtain:
\begin{equation}
*dB^{\ivec }= \Omega^{\ivec \jvec } \frac{1}{\sqrt{|g|}} \partial_\nu \left( \sqrt{|g|} M_{\jvec \kvec }  F^{\nu \mu \kvec }\right) g_{\mu \rho} dx^\rho \,,
\end{equation}
so we can use the one-form field equation \eqref{SU4OneFormFieldEquation} to find:
\begin{equation}
\begin{split}
*dB^{\ivec } & = - \tfrac{1}{12} \Omega^{\ivec \jvec } M^{CD} D_\mu M_{BD} \left( 
X_{\jvec C}{}^B - 24 K_{\jvec }{}^B \theta_C 
\right)  dx^\mu \\
& = 
\begin{pmatrix}
0\\
0\\
2\sqrt{3} \frac{e^{-12 U}}{R} \left( Da -\frac{3 i}{4} \left( {\chi}^* D{\chi} - {\chi} D{\chi}^* \right) \right)\\
\\
6\frac{ e^{-12 U}}{R} ( \left(|\chi|^2 - 1\right) \left( Da -\frac{3 i}{4} \left( {\chi}^* D{\chi} - {\chi} D{\chi}^* \right) \right)- i  e^{6 U} \left( {\chi}^* D{\chi} - {\chi} D{\chi}^*  \right)	)
\end{pmatrix}
\,.
\end{split}
\end{equation}
We see from this that $dB_{(1)} = H_3^{\text{there}}$ gives the three-form field defined in (2.7) of \cite{Gauntlett:2009bh}.

\vspace{1em}
\noindent \emph{Scalar equation.}
Finally, the scalar equation of motion is reproduced by \eqref{ScalarEom_CT}. In particular, if we demand the scalars to take constant values, it takes the form $ V_{AB}=0\,,$ which reproduces the extremisation of the effective potential \eqref{VeffSU4}. The contribution of all the components of the fluxes is crucial to obtain this result, in the same way it was crucial to fully determine the effective potential \eqref{VeffSU4}.

\subsection{Example: finite and infinite truncations with $G_{\text{sym}}=\mathrm{SO}(7)$} \label{sec:ExampleSO7}

For our second example, we take $\Gsym=\mathrm{SO}(7)$.
There are three inequivalent embeddings of $\mathrm{SO}(7) \subset \mathrm{SO}(8)$, due to triality.
Each of these embeddings branches only one of the three eight-dimensional representations of $\mathrm{SO}(8)$ to the $\mathbf{7} + \mathbf{1}$ of $\mathrm{SO}(7)$, with the other two being irreducible.  We will denote these three embeddings by $\mathrm{SO}(7)_v$, $\mathrm{SO}(7)_s$ and $\mathrm{SO}(7)_c$, where the subindex indicates which $8$-dimensional representation of $\mathrm{SO}(8)$ is branched to the $\mathbf{7} + \mathbf{1}$. In what follows we discuss these three cases, two of them leading to a consistent truncation first worked out in \cite{Gauntlett:2009zw} and the remaining one going beyond a standard finite consistent truncation and providing an example of an infinite consistent truncation.

\paragraph{The homogeneous cases: $\Gsym = \mathrm{SO}(7)_{c,s}$}
Let's first discuss the case $\Gsym = \mathrm{SO}(7)_c$ symmetry. 
The group $\mathrm{SO}(7)_c$ acts homogeneously on $S^7$, as $S^7 \cong \mathrm{SO}(7)_c/G_2$. This results in $\Mint / ( \Gsym/H )$ being trivial, which implies that, as in the $\mathrm{SU}(4)_c$ case, $\mathcal{V}$ is independent of internal coordinates. Moreover, $L(y)$ is given by the coset $ \mathrm{SO}(7)_c/G_2$. Instead of working out this example explicitly from the beginning, we note that by means of the chain of isomorphisms $\mathrm{SU}(4)_c / \mathrm{SU}(3) \cong S^7 \cong \mathrm{SO}(7)_c/G_2$ we can directly study $\mathrm{SO}(7)_c$ symmetric configurations as a restriction of the $\mathrm{SU}(4)_c$ symmetric ones, the only step required being the proper truncation of the field content with respect to the generalised $G_2$ structure imposed by the weak $G_2$ truncation, exactly along the lines of \cite{Gauntlett:2009zw}. In terms of the scalar fields, we find that the identifications:
\begin{equation}
U =V \,, \quad \chi_2 = \tfrac{2}{\sqrt{3}} h \,, \quad \chi_1 = 0 \,, \quad a = 0 \,,
\end{equation}
guarantee that $\mathcal{V}(x)$ lies in  $\mathcal{C}_{\Edd}(G_2)$. Moreover, there are no $G_2$ invariant vector fields, so the resulting $\mathcal{N}=1$ supergravity does not have any vector fields.
It is straightforward to check that starting from the $\mathrm{SU}(4)_c$ symmetric case of the previous subsection, these identifications lead to the weak $G_2$ supergravity of \cite{Gauntlett:2009zw}. 
The case $\Gsym = \mathrm{SO}(7)_s$ similarly follows from the analogous $\Gsym = \mathrm{SU}(4)_s$ case.

\paragraph{The inhomogeneous case: $\Gsym = \mathrm{SO}(7)_{v}$}

As we discussed in section \ref{SingletTowers}, the action of $\mathrm{SO}(7)_v$ on $S^7$ is not homogeneous. 
The singlet solution again has the form \eqref{SingletSolutionExponential}.
The coset element is $L(y) \in \mathrm{SO}(7)_v/\mathrm{SO}(6)_v$, we have $\mathcal{V} = e^{-b}$ where algebraically $b$ lies in $\mathcal{C}_{\Edd}(\mathrm{SO}(6))$ and may depend both on the external coordinates $x$ as well as the space $\mathrm{SO}(7)/\mathrm{SO}(6)$ which is not trivial.

\begin{table}[h]
\centering
\renewcommand{\arraystretch}{1.2}

\begin{tabular}{|c|c|c|}
\hline
\multicolumn{3}{|c|}{$\Gsym = \mathrm{SO}(7)_v$ \hspace{1em} $H= \mathrm{SO}(6)_v$} \\ \hline
Parallelisation & $ U_A = \mathring{U}_B (L^{-1})_A{}^B$ &   \\ & $L \in \mathrm{SO}(7)_v/ \mathrm{SO}(6)_v$ &  \eqref{SO7vL(y)8v}\\\hline
Scalars & $\mathcal{C}_{\Gseven}(\mathrm{SO}(6)_v) = \mathrm{SL}(2) \times \mathrm{SO}(1,1)$ & \\
 & $\mathcal{V}(x,\theta_7) \in \tfrac{ \mathrm{SL}(2) \times \mathrm{SO}(1,1)}{\mathrm{SO}(2)}$ & \eqref{SO7cosetparam}\\\hline
Vectors & 2 invariant $K_{\ivec }$  &   \eqref{SO6InvariantVectorsExplicit}\\ 
 & $A_\mu = K_{\ivec } A_\mu{}^{\ivec }$ & \\ \hline
2-forms & 4 invariant $J^{\aadj }$ & Generators of $\mathrm{SL}(2) \times \mathrm{SO}(1,1)$ \\
 & $B_{\mu\nu} = J^{\aadj } B_{\mu\nu \aadj }$ & 
\\ \hline
\end{tabular}

\caption{Ingredients for the $\mathrm{SO}(7)_v$ symmetric deformation of the $S^7$ generalised parallelisation. This leads to an infinite-mode consistent truncation.} 
\label{so7stuff} 
\end{table}

To evaluate the equations of motion for this case, it's convenient to use the following form of the deformed generalised parallelisation: 
\begin{equation}
\gM_{MN}  = E_M{}^A (x,y) E_M{}^B (x,y) \delta_{AB} \,,\quad
E^A{}_M = \Ur_M{}^C L_C{}^B(y)\mathcal{V}_B{}^A(x,y)
\label{SO7vConfiguration}
\end{equation}
and to use the equations of motion as presented in appendix \ref{EomGenDef} (with $\Mflat_{AB} = \delta_{AB}$).
The non-trivial internal coordinate dependence of $\mathcal{V}_B{}^A$ means that this model gives rise to an infinite consistent truncation. For instance, if we compute the effective potential \eqref{Veff} --  assuming here the four-dimensional external metric is independent of $\theta_7$ -- we find
\begin{equation}
\begin{split}
V_{\text{eff}}= \frac{e^{-3 \phi (\theta_7)}}{2 R^2} \Big(&4 \chi (\theta_7) \varphi '(\theta_7)+\varphi '(\theta_7)^2+12 \cot (\theta_7) \varphi '(\theta_7)\\
& +e^{2 \varphi (\theta_7)} \left(\chi '(\theta_7)-\chi (\theta_7)^2-6 \chi (\theta_7) \cot (\theta_7)+5\right)^2-\sinh (2 \varphi (\theta_7))\\
& +\cosh (2 \varphi (\theta_7)) +2 \chi '(\theta_7)+2 \chi (\theta_7)^2+12 \chi (\theta_7) \cot (\theta_7)-6 \phi ''(\theta_7)\\
& +12 \phi '(\theta_7)^2-60 \cot (\theta_7) \phi '(\theta_7)-60 \left(\chi (\theta_7) \cot (\theta_7)-1\right)^2 e^{\varphi (\theta_7)+2 \phi (\theta_7)}\\
& -60 \cot ^2(\theta_7) e^{2 \phi (\theta_7)-\varphi (\theta_7)}+60 \csc ^2(\theta_7)-74\Big)\, .
\end{split}
\label{SO7vVeff}
\end{equation}
This potential \eqref{SO7vVeff} still depends on $\theta_7$, which is a coordinate on $S^7$ defined through \eqref{SO7vEmbeddingCoordinates}.

We can understand this feature as follows. For homogeneous cases, the branching of the $\mathbf{8}_v$ of $\mathrm{SO}(8)$ under $G_{\text{sym}}$ does not provide singlets (nor does any symmetrised tensor product $(\mathbf{8}_v \otimes \mathbf{8}_v \otimes ...)_s$). This implies that none of the embedding coordinates of the sphere, which span the $\mathbf{8}_v$ of $\mathrm{SO}(8)$, can appear in the effective potential, as it must be a singlet under $G_{\text{sym}}$. However, when the action of $G_{\text{sym}}$ is inhomogeneous, there is at least one fixed point in the sphere under the action of $G_{\text{sym}}$. This fixed point is a singlet of $G_{\text{sym}}$ and therefore it can appear in the potential. In the current example, using the parametrisation \eqref{SO7vEmbeddingCoordinates}, we find that $y_8=\cos (\theta_7)$ is a singlet of $\mathrm{SO}(7)_v$. 
More generally, any function $f$ obeying $\tfrac12 a{}^{\as \bs} L_{v_{\as \bs}} f=0$ for all the $\mathrm{SO}(7)_v$ generators is a singlet, which for this parametrisation allows $f=f(\theta_7)$.

We should then consider the remaining fields in this infinite-mode consistent truncation.
For instance, this configuration has two invariant vectors, arising as the two $\mathrm{SO}(6)_v$ singlets in the $\mathbf{56}$ of $\Gseven$.
These vector fields as well as the external metric should in principle be allowed to depend on $\theta_7$, and the equations of motion of section \ref{EomGenDef} used to search for solutions. 
However, as a preliminary analysis of this example of an infinite consistent truncation, we can search for `vacuum' solutions, for which the gauge fields vanish, the metric depends only on the four-dimensional external coordinates, and	 the scalars depend only on $\theta_7$ and are constant with respect to the external spacetime. Then all we have to solve are the Einstein equation of the form $0 =R_{\mu \nu}(\bar g) - \tfrac{1}{2} \bar g_{\mu\nu} \left( R(\bar g) - V_{\text{eff}}\right)$ together with the scalar field equation \eqref{ScalarEom_Ansatzed} which reduces with these assumptions to the vanishing of $V_{AB}$ of \eqref{VAB}.
This leads to only three independent equations which we record in the appendix as \eqref{lovely1}, \eqref{lovely2} and \eqref{lovely3}.

It can be checked that for constant scalar fields these equations are satisfied for the values $e^{2\varphi}=e^{4\phi}=5$, $\chi=0$, for which the potential evaluates to $V_{\text{eff}}=-\frac{8\cdot 5^{3/4}}{R^2}$,
recovering in this language the $\mathrm{SO}(7)_v$ solution of \cite{deWit:1984va} in the conventions of  \cite{Larios:2019kbw}.
This result has an easy underlying explanation. The  values of scalars leading to the solution enhance the $\mathrm{SO}(6)_v$-structure to a $\mathrm{SO}(7)_v$-structure, and hence $\mathcal{V}(x)$ commutes with $\mathrm{SO}(7)_v$. Consequently, it follows that $L(y) \mathcal{V}(x)=\mathcal{V}(x) L(y)$, so that the $L(y)$ deformation becomes trivial and we are back in the usual case of a finite consistent truncation to maximal supergravity.

It would be interesting to scan for numerical solutions  of \eqref{lovely1}--\eqref{lovely3} involving $\theta_7$-dependent scalars. Genuine vacuum solutions with AdS (or Minkowski) four-dimensional spacetime would require that the scalars give a constant $V_{\text{eff}}$ in the Einstein equation, which is far from guaranteed. More generally, one could consider an appropriate ansatz involving a $\theta_7$-dependent four-dimensional spacetime and look for solutions of this form, which would still be solutions of the full 11-dimensional supergravity.

\section{Discussion} \label{sec:discussion}

Let us recap. 
We start with a background $\Mint$ which admits a generalised parallelisation, with generalised metric factorising $\gM_{MN}(x,y) = \Ur_M{}^A (y)\Ur_N{}^B(y)\Mr_{AB}(x)$, leading to a consistent truncation with gauge group $\Ggauging$.
We assume that $\Mr_{AB}$ is invariant under $\Ggauging$, such that $\Ggauging$ corresponds to isometries of $\Mint$.
We select some subgroup $\Gsym \subseteq \Ggauging \subset \Edd$.
Singlets under this subgroup are functions valued in the Lie algebra of the global $\Edd$ determined by the parallelisation, and are defined as solutions to the equation \eqref{SingletEquation}.
Using the action of $\Gsym$ on $\Mint$, we can find solutions of this equation of the form \eqref{SingletSolution}.
This uses as data i) the orbits on $\Mint$ of $\Gsym$, which are cosets $\Gsym/H$ for some subgroup $H$, and ii) the commutant of this subgroup $H$ in $\Edd$.
After exponentiating these singlets we can build globally defined `deformations' of the original generalised parallelisation, and use these to define new consistent truncations.

The consistent truncations associated to the corresponding deformed generalised parallelisation retain a finite or infinite set of modes depending on whether the action of $\Gsym$ on $\Mint$ is homogeneous or not. In all cases, the uplifts to the conventional formulation of supergravity still follow from the usual generalised Scherk–Schwarz prescription \cite{Lee:2014mla, Hohm:2014qga,Varela:2015ywx}. For example, the uplift of SO(8)-gauged supergravity on $S^7$ to the $D=11$ metric takes on the explicit form \cite{Varela:2015ywx}
\begin{equation} \label{KKEmbedding}
d\hat{s}_{11}^2 =  \Delta^{-1} \, ds_4^2  \, +\tfrac{1}{12}  \, \Delta^2 \, (t_{\cal A}{}^{\cal B} )_A{}^C \, (t_{\cal C}{}^{\cal D} )_B{}^D \, {\bar M}^{AB}   \, {\bar M}_{CD}  \, \mu_{\cal B} \mu_{\cal D}  D \mu^{\cal A} D\mu^{\cal C}      \; , 
\end{equation}
with warp factor, $\Delta$, and covariant derivatives, $D$, specified in \cite{Varela:2015ywx}. Both type of truncations, finite and infinite, discussed in this paper still uplift as in ({\ref{KKEmbedding}), with the scalar matrix of the form ${\bar M}_{AB} = (P^{-1})_A{}^C \delta_{CD} (P^{-1})_B{}^D$ either $\mu^A$-independent or dependent, respectively. In the latter case, the $\mu^{\cal A}$ dependence of ${\bar M}_{AB}$ is constrained in such a way that $\Gsym$ is contained in the isometry group of (\ref{KKEmbedding}). 

We have presented some examples leading to known finite consistent truncations. The finite consistent truncation of $D=11$ supergravity on the squashed $S^7$ of \cite{Awada:1982pk} presented in \cite{Page:1983mke} was obtained using these methods with $\Gsym = \mathrm{Sp}(2) \times \mathrm{Sp}(1)$ in \cite{Duboeuf:2022mam,Duboeuf:2023dmq}. This finite truncation on the squashed $S^7$ can be enlarged to a full $D=4$ ${\cal N}=4$ supergravity \cite{Cassani:2011fu}. The latter can be recovered using our formalism with $\Gsym = \mathrm{Sp}(2)$, and associated generalised Sp(1) structure in the language of \cite{Cassani:2019vcl,Cassani:2020cod}.

More interestingly, we have provided a formalism to obtain consistent truncations to infinite sets of KK fields, and we have specified the circumstances under which those can be obtained. We have illustrated this method with a new infinite consistent truncation associated to the inhomogeneous action of SO$(7)_v$ on $S^7$. This example admits further generalisations for $\Gsym = \mathrm{SO}(d-1) \subset \mathrm{SO}(d) \subset \mathrm{SL}(d)$ and general $d$. There is a uniform description of consistent truncations with $\Ggauging  = \mathrm{SO}(d)$ using a generalised parallelisation of the sphere $S^{d-1}$ in $\mathrm{SL}(d)$ generalised geometry \cite{Lee:2014mla}.
For $d=4$ this corresponds to a generalised parallelisation on $S^3$ using the $\mathrm{SL}(4) \sim \mathrm{SO}(3,3)$ generalised geometric description of the NSNS sector in 10-dimensional supergravity. For $d=5$ this corresponds to the consistent truncation of 11-dimensional supergravity on $S^4$, using $\Gfour	$ exceptional generalised geometry.
For $d=6$, this gives the consistent truncation of 10-dimensional type IIB supergravity on $S^5$, using an $\mathrm{SL}(6)$ subgroup of $\Gsix$ generalised geometry \cite{Baguet:2015sma}.
For $d=7$, we have the $S^7$ in $\Gseven$ generalised geometry that we have focused on in this paper.
In all these cases, the choice $\Gsym = \mathrm{SO}(d-1)$ leads to an inhomogeneous action, an $\mathrm{SO}(d-2)$ structure and a commutant $\mathcal{C}_{\mathrm{SL}(d)} (\mathrm{SO}(d-2)) = \mathrm{SL}(2) \times \mathrm{SO}(1,1)$. The corresponding `scalars' then have non-trivial internal coordinate dependence.
We have for example analysed the equations of motion following from the $d=4$ case and found similar expressions to \eqref{lovely1} to \eqref{lovely3}.

Another example worth considering would be $\Gsym = \mathrm{SU}(3)$ in $S^7$, leading to an associated generalised SU(2)-structure. This would again lead to an infinite consistent truncation with $\mathcal{C}_{\Gseven} (\mathrm{SU}(2)) = \mathrm{SL}(2) \times \mathrm{SO}(6,3)$, and $S^7$ dependence on the corresponding `scalars'. The equations of motion of appendix \ref{EomGenDef} particularised to this setup must contain at least two interesting anti-de Sitter solutions with residual ${\cal N}=2$ supersymmetry and $\mathrm{SU}(3) \times \mathrm{U}(1)$ bosonic symmetry: those first obtained in \cite{Corrado:2001nv} and \cite{Gabella:2012rc}. The former can in fact be obtained as a solution of the finite truncation associated to the conventional parallelisation on $S^7$, namely, as a vacuum of SO(8)-gauged supergravity. The solution of \cite{Gabella:2012rc}, however, lies outside the gauged supergravity truncation and has naturally associated an infinite-dimensional truncation of the type we have described in this paper. While the solution of \cite{Gabella:2012rc} has already been briefly described within EGG \cite{Ashmore:2016qvs}, it would be very interesting to recover it within our formalism, in order to open up scope for further developments.

The existence of these already known solutions to our field equations cries for a systematic search of supergravity solutions using our formalism.

\section*{Acknowledgements} 
We would like to thank Mattia Ces\`aro for collaboration at the initial stage of this project. MP is supported by predoctoral award FPU22/02084 from the Spanish Government. OV is supported by NSF grant PHY-2310223.

This work is also supported through the grants CEX2020-001007-S and PID2021-123017NB-I00, funded by MCIN/AEI/10.13039/501100011033 and by ERDF A way of making Europe.

\appendix

\section{Equations of motion for deformed generalised parallelisations} 
\label{eom}

%%%%%%%%%%%%%%%%%%%%%%%%%%%%%%%%%%%%%%%%%%%%%%%%%%%%%%%%%%%%%%%%%%%%%%%%
\subsection{Supergravity equations of motion from exceptional field theory} \label{sec:Eoms}

The action of supergravity in the full exceptional field theory approach \cite{Hohm:2013pua, Berman:2020tqn} is
\be
\begin{split}
S & = \int \dd^n x \, \dd y \,\sqrt{|g|} \big( \hat R(g) 
+\tfrac{1}{4\alpha}  \mathcal{D}_\mu \gM^{MN} \mathcal{D}^\mu \gM_{MN} 
- V(\gM, g)
\\ & \qquad\qquad\qquad\qquad - \tfrac{c_A}{4}\gM_{MN} \Fa_{\mu\nu}{}^M \Fa^{\mu\nu}{}^N + \ldots 
\big) + \int \mathcal{L}_{CS} \,.
\end{split} 
\label{SExFT}
\ee
While we are mostly interested in the case $n=4$, described by the $\Gseven$ ExFT \cite{Hohm:2013uia}, we keep our presentation general where possible.
Thus, the dots in \eqref{SExFT} denote further gauge field kinetic terms: these are not present for $n=4$, and we will ignore them below.
The numerical constant $\alpha$ is that appearing in the generalised Lie derivative \eqref{genLie}, and for $\Gseven$ is given by $\alpha=12$.
The coefficient $c_A$  of the kinetic term for the one-form gauge field is equal to $1/2$ for $\Gseven$, and is $1$ otherwise.
This reflects the fact in the $\Gseven$ case, the field strength of this field is subject to a self-duality constraint imposed by hand: 
\be
\mathcal{F}_{\mu\nu}{}^M = - \tfrac12 |g|^{1/2} \epsilon_{\mu\nu\rho\sigma} \Omega^{MN} \gM_{NK} \mathcal{F}^{\rho\sigma K} \,.
\label{SelfDuality}
\ee
The `scalar potential' is: 
\be
\begin{split}
V =&  - \tfrac{1}{4 \alpha} \gM^{MN} \partial_M \gM^{KL} \partial_N \gM_{KL} + \tfrac{1}{2} \gM^{MN} \partial_M \gM^{KL} \partial_L \gM_{NK} \\
& - \tfrac{1}{2} \partial_M \ln |g| \partial_N \gM^{MN} 
- \tfrac{1}{4} \gM^{MN} \partial_M \ln |g| \partial_N \ln |g|
- \tfrac{1}{4} \gM^{MN} \partial_M g^{\mu \nu} \partial_N g_{\mu\nu}\,.
\label{VExFT}
\end{split} 
\ee
Finally there is a Chern-Simons term whose definition is dimension dependent.
For $\Gseven$, this is defined such that the variation equals \eqref{ChernSimonsContribs}, below.
The equations of motion following from varying the ExFT action \eqref{SExFT} are then as follows.

\noindent \emph{Einstein equation.}
Varying \eqref{SExFT} with respect to $g^{\mu\nu}$ we obtain the following Einstein equation:
\be
\begin{split}
0 & = 
\hat R_{\mu \nu} - \tfrac{1}{2} g_{\mu\nu} \left( \hat R[g] + \tfrac{1}{4 \alpha}  g^{\rho\sigma} \mathcal{D}_\rho \mathcal{M}_{MN} \mathcal{D}_\sigma \mathcal{M}^{MN} - \tfrac{c_A}{4} \gM_{MN} \Fa_{\rho \sigma}{}^M \Fa^{\rho \sigma N} \right) 
\\ & \qquad
+ V_{\mu\nu} 
+ \tfrac{1}{4 \alpha} \mathcal{D}_\mu \gM_{MN} \mathcal{D}_\nu \gM^{MN} 
- \tfrac{c_A}{2 } \gM_{MN} \Fa_{\mu \rho}{}^M \Fa_{\nu \sigma}{}^N g^{\rho \sigma}
 \,,
\end{split} 
\label{ExcEin} 
\ee 
where we have collected the following non-manifestly covariant terms involving internal derivatives:
\be
\begin{split} 
V_{\mu\nu}  &\equiv  \tfrac12 g_{\mu\nu} V( \mathcal{M},g )
+ \tfrac{1}{2}|g|^{-1/2} g_{\mu\nu} \partial_M \left( |g|^{1/2} ( \partial_N \gM^{MN} + \gM^{MN} \partial_N \ln |g| )\right) 
\\ & \quad
- \tfrac{1}{2} |g|^{-1/2}\partial_M ( |g|^{1/2} \gM^{MN} ) \partial_N g_{\mu\nu}  
- \tfrac{1}{2} \gM^{MN} g_{\mu \rho} \partial_M g^{\rho \sigma} \partial_N g_{\sigma \nu} 
- \tfrac{1}{2} \gM^{MN} \partial_M \partial_N g_{\mu\nu}\,.
\label{Veff_Ein}
\end{split}
\ee
The Ricci tensor and scalar are given by the usual expressions
\be
\hat R_{\mu\nu} = \mathcal{D}_\rho \Gamma_{\mu\nu}{}^\rho - \mathcal{D}_\mu \Gamma_{\nu \rho}{}^\rho 
+ \Gamma_{\rho\lambda}{}^\rho \Gamma_{\mu\nu}{}^\lambda 
- \Gamma_{\nu\lambda}{}^\rho\Gamma_{\rho \mu}{}^\lambda 
\,,\quad
\hat R = g^{\mu\nu} \hat R_{\mu\nu} \,.
\label{ExceptionalRicci}
\ee
in terms of the standard Christoffel symbols, where however all derivatives are covariantised under generalised diffeomorphisms via $\mathcal{D}_\mu = \partial_\mu - \mathcal{L}_{A_\mu}$. 

\noindent \emph{Scalar equation.}
Varying the action \eqref{SExFT} with respect to $\gM^{MN}$, and taking into account that $\gM$ parametrises a coset, we obtain what we refer to as the scalar equation of motion:
\begin{subequations}
\begin{align}
0 &  = \gM_{P(M} \mathbb{P}^P{}_{N)}{}^K{}_Q \gM^{LQ} \mathcal{K}_{KL} \,,
\label{scalarEom1}\\
\mathcal{K}_{MN} & = 
- \tfrac{1}{4\alpha}|g|^{-1/2}  \mathcal{D}_\mu( |g|^{1/2} \mathcal{D}^\mu \gM_{MN} ) 
+ \tfrac{1}{4\alpha}|g|^{-1/2} \gM_{MK} \gM_{NL}  \mathcal{D}_\mu( |g|^{1/2} \mathcal{D}^\mu \gM^{KL} ) \notag
\\ & + \tfrac{c_A}{2} \gM_{MK} \gM_{NL} \mathcal{F}_{\mu\nu}{}^K \mathcal{F}^{\mu\nu}{}_L 
+ V_{(MN)}\,,
\label{scalarEom2} 
\end{align}
\label{scalarEom}
\end{subequations} 
where we again collect the terms involving bare $\partial_M$ derivatives:
\be\begin{split}
V_{MN} & =  \tfrac{1}{4\alpha} \partial_M \gM^{KL} \partial_N \gM_{KL} - \tfrac12 \partial_M \gM^{KL} \partial_L \gM_{NK} 
\\ & 
- \tfrac{1}{4\alpha} \partial_K ( \gM^{KL}[ \partial_L \gM_{MN} - 2 \alpha \partial_M \gM_{LN} ) 
\\ &
+ \tfrac{1}{4\alpha} \gM_{KM} \gM_{LN} \partial_P ( \gM^{PQ} \partial_Q \gM^{KL} - 2 \alpha \gM^{KQ} \partial_Q \gM^{LP} )
\\ & 
- \tfrac{1}{4\alpha} \partial_P \ln |g| \gM^{PQ} ( \partial_Q \gM_{MN} - 2 \alpha \partial_M \gM_{QN} ) 
+ \tfrac14 \partial_M g^{\mu\nu} \partial_N g_{\mu\nu} - \tfrac12 \partial_M \partial_N \ln |g| \,.
\end{split}
\label{VMN}
\ee
\noindent \emph{Gauge field equation.}
Finally, varying the action \eqref{SExFT} with respect to $\Aa_\mu^{M}$ gives:
\be
\begin{split}
0 & = 
c_A
|g|^{-1/2} \mathcal{D}_\nu \left[|g|^{1/2} \gM_{MN} \Fa^{\nu\mu N} \right]+ |g|^{-1/2} \frac{\delta \mathcal{L}_{\text{CS}}}{\delta \Aa_\mu{}^M}
 \\ & \quad
 -\tfrac{1}{2\alpha}  \partial_M \gM_{KL} \mathcal{D}^\mu \gM^{KL} 
 + |g|^{-1/2} \partial_P ( |g|^{1/2} \gM_{KM} \mathcal{D}^\mu \gM^{PK} )
\\ & \quad
 - \tfrac{1}{2} g^{\mu\lambda} \partial_M g^{\nu\rho} \mathcal{D}_\lambda g_{\nu\rho} 
+g^{\mu\lambda} \partial_Mg^{\nu\rho} \mathcal{D}_\nu g_{\rho\lambda} 
+\tfrac{1}{2} \partial_M g^{\mu\nu} \mathcal{D}_\nu\ln |g|
\\ & \quad
 +g^{\mu\nu} \partial_M \mathcal{D}_\nu \ln |g| + \partial_M \mathcal{D}_\nu g^{\mu\nu} \,.
\end{split}
\label{Aeom} 
\ee
For $\Gseven$, $c_A =1/2$, and the Chern-Simons contributions are \cite{Hohm:2013uia}
\be
|g|^{-1/2}
\frac{\delta \mathcal{L}_{\text{CS}}}{\delta \Aa_\mu{}^M} = 
 - \tfrac14 |g|^{-1/2} \epsilon^{\mu\nu\rho\sigma} \mathcal{D}_\nu \mathcal{F}_{\rho\sigma}{}^N \Omega_{NM}\,,
 \label{ChernSimonsContribs}
\ee
which double-up with the first term in \eqref{Aeom} on using the self-duality condition \eqref{SelfDuality}.

\subsection{Equations of motion for deformed generalised parallelisations} 
\label{EomGenDef}

We now rewrite the above equations of motion in a way that is adapted to describe deformations of generalised parallelisations.
We will begin with the following ansatz
\be
g_{\mu\nu}(x,y) = \Delta^2(y) \bar g_{\mu\nu}(x,y)\,,\quad
\gM_{MN}(x,y) = E_M{}^A (x,y) \Mflat_{AB} E_N{}^B(x,y) \,,
\label{Efactor}
\ee
where $\Mflat_{AB}$ is assumed constant.
For $\gM_{MN}$, this is completely general and always possible locally. 
Similarly, contrary to the conventional ExFT generalised Scherk-Schwarz ansatz \cite{Berman:2012uy,Hohm:2014qga}, we allow for a possible internal coordinate dependence of the external metric, subject to the factorisation involving the weighted scalar $\Delta(y)$. The above can be viewed as the starting point for a `flux formulation' of ExFT, for the internal sector see \cite{Aldazabal:2013mya, Blair:2014zba} and a version for the full theory in the $\Gfour$ case see \cite{Gubarev:2023kvq}.

Treating $\Delta$ as a scalar of weight $-\omega = \tfrac{1}{n-2}$, we define a weighted generalised frame 
$\Ev^M{}_A = \Delta E^M{}_A$.
We then define `generalised fluxes' aka \emph{non-constant} trombone and embedding tensor gaugings via the usual ExFT formulae:
\be
\mathcal{L}_{\Ev_A} \Delta = \theta_A \Delta\,,\quad
\mathcal{L}_{\Ev_A} \Ev_B = - X_{AB}{}^C \Ev_C \,,
\ee
where the embedding tensor $\Theta_{AB}{}^C$ and trombone $\theta_A$ are defined via:
\begin{subequations}
\label{defGaugings}
\begin{align}
X_{AB}{}^C &= \Theta_{AB}{}^C - \delta^C_B \theta_A + \alpha \tfrac{n-2}{n-1} \mathbb{P}^C{}_B{}^D{}_A \theta_D \,,
\label{defX}
\\
\theta_A & = \tfrac{1}{n-2} \Delta ( (n-1) E^N{}_A \partial_N \ln \Delta - \Omega_{BA}{}^B) \,,
\label{deftheta} 
\\
\Theta_{AB}{}^C & = \Delta \big(
\Omega_{AB}{}^C - \alpha \mathbb{P}^C{}_B{}^D{}_E \Omega_{DA}{}^E  + \tfrac{\alpha}{n-1} \mathbb{P}^C{}_B{}^D{}_A \Omega_{ED}{}^E 
\big) \,,
\label{defTheta}
\end{align} 
\end{subequations} 
where $\Omega_{AB}{}^C \equiv E^M{}_A E^N{}_B \partial_M E_N{}^C$ and $\mathbb{P}^A{}_B{}^C{}_D$ is the adjoint projector. The former is adjoint projected: $\mathbb{P}^A{}_B{}^C{}_D \Omega_{EC}{}^D =\Omega_{EB}{}^A$.
The definition \eqref{defTheta} of the embedding tensor can be written as 
\be
\Theta_{AB}{}^C = \kappa (P_{R_\Theta})_{AB}{}^C,{}^{DE}{}_{F} \Delta \Omega_{DE}{}^F
\ee
for some projector onto the embedding tensor representation (denoted $R_{\Theta}$ here) and a constant $\kappa$.
For $\Gseven$ in particular, $R_{\Theta} = {\bf 912}$ and $\kappa=7$ (see appendix \ref{app_7rep}).

Covariance under generalised diffeomorphisms guarantees that all terms in the equations of motion will regroup in terms of the above generalised fluxes.
This covariance is manifest for all terms excluding the contributions $V_{\mu\nu}$ and $V_{MN}$ to the Einstein and scalar equations of motion.
After some calculation, the internal contribution $V_{\mu\nu}$, given in \eqref{Veff_Ein}, to the Einstein equation \eqref{ExcEin}, can be shown to take the form
\be
\begin{split}
V_{\mu\nu} & = \tfrac12 \bar g_{\mu\nu}  V_{\text{eff}}
- \Mflat^{AB} \big(
\tfrac14 \hat\partial_A \ln |\bar g| \,\hat\partial_B \bar g_{\mu\nu} 
+ \tfrac12 \bar g_{\mu\rho} \hat\partial_A \bar g^{\rho \sigma} \hat\partial_B \bar g_{\sigma \nu} 
\\ & \qquad\qquad\qquad\qquad\quad
+ \tfrac12 \hat\partial_A \hat\partial_B \bar g_{\mu\nu} 
+ \tfrac{n-2}{2} \theta_A \hat\partial_B \bar g_{\mu\nu} 
\big) \,,
\end{split}
\ee
where we have defined $\hat \partial_A = \Delta E^M{}_A \partial_M$,
and we have an `effective potential'
\be
\begin{split}	
V_{\text{eff}} & \equiv     \tfrac{1}{2\alpha \kappa}(  \Mflat^{AB} \Mflat^{CD} \Mflat_{EF} \Theta_{AC}{}^E \Theta_{BD}{}^F 
 +\kappa \Mflat^{AB} \Theta_{AC}{}^D \Theta_{BD}{}^C ) %\notag
 \\  & \quad  +  \tfrac{(n-2)^3}{n-1}  \Mflat^{AB} \theta_A \theta_B
  + 2(n-2)\Mflat^{AB} \hat{\partial}_A \theta_B  
\\ & \quad
+ \Mflat^{AB} \left(
\tfrac14 ( \hat\partial_A \ln |\bar g| \hat \partial_B \ln |\bar g| - \hat\partial_A \bar g_{\mu\nu} \hat\partial_B \bar g^{\mu\nu}) + \hat \partial_A \hat \partial_B \ln| \bar g|
+ (n-2) \theta_A \hat \partial_B \ln| g |
\right) \,.
\end{split}
\label{Veff}
\ee
Indeed, when $\theta_A=0$ and $\hat\partial_A \bar g_{\mu\nu}=0$, this reproduces the form of the scalar potential of maximal gauged supergravity. (Note that for $n=7$, with $\mathrm{E}_{4(4)}= \Gfour$, $R_{\Theta}$ is a sum of two irreducible representations, leading to a slightly different structure.)
Meanwhile, the internal contribution $V_{MN}$, given in \eqref{VMN}, to the scalar equation \eqref{scalarEom}, factorises as $\gM_{P(M} \mathbb{P}^P{}_{N)}{}^K{}_Q \gM^{QL} {V}_{KL} = E_M{}^A E_N{}^B V_{AB}$ 
with
\begin{align}
- V_{AB} &  = \notag
\Mflat^{CD} \Mflat_{E(A|} \hat{\partial}_C \Theta_{D|B)}{}^E 
+  \alpha\tfrac{n(n-2)}{(n-1)}  \Mflat_{E(A} \mathbb{P}^E{}_{B)}{}^C{}_F \Mflat^{FD} \hat{\partial}_C \theta_D 
\\ & \quad \notag
+ (n-2)   \theta_C \Mflat_{E(A} \Mflat^{CD} \Theta_{|D|B)}{}^E\\
& \quad+\tfrac{1}{2 \kappa} \Mflat_{A'(A} \mathbb{P}^{A'}{}_{B)}{}^C{}_D \Mflat^{DD'} \notag
   \Big( \Mflat^{EF} \Mflat_{GH} ( \Theta_{CE}{}^G \Theta_{DF}{}^H+ \Theta_{EC}{}^G \Theta_{FD}{}^H)
 %\\ & \qquad  \quad\quad \quad \quad \quad \quad \quad \quad \quad \quad  
\\ & \qquad \qquad\qquad\qquad\qquad\qquad\quad\notag  + \kappa \Theta_{CE}{}^F \Theta_{DF}{}^E - \Mflat^{EF} \Mflat^{GH} \Mflat_{CG'} \Mflat_{DH'} \Theta_{EG}{}^{G'} \Theta_{FH}{}^{H'} 
\Big)\label{VAB}
\\ & \notag
\quad
+ \Mflat_{C(B} \mathbb{P}^C{}_{A)}{}^{E}{}_D \Mflat^{D F} \left(
- \tfrac14 \hat\partial_E \bar g_{\mu\nu} \hat\partial_F \bar g^{\mu\nu} 
+ \tfrac12 \hat \partial_E \hat \partial_F \ln |\bar g|
+ \tfrac12 \tfrac{n-2}{n-1} \theta_E \hat \partial_F \ln |\bar g|
\right) 
\\ & 
\quad
+ \tfrac{1}{2\alpha} \hat \partial_C \ln |\bar g| \Mflat^{CD} \Theta_{D(A}{}^E \Mflat_{B)E} \,.
\end{align}
To obtain the full equations of motion, we also need to specify an ansatz for the $p$-form gauge fields.
We again allow an expansion in terms of $p$-forms depending both on $x$ and $y$, so that:
\be
\begin{split}
\Aa_\mu{}^M(x,y) = \Delta(y) E^M{}_A(x,y)  \Aflat_\mu{}^A (x,y) \,,\quad
\Ab_{\mu\nu \alpha}(x,y) = \Delta^2(y) E_\alpha{}^\beta(x,y)  \Bflat_{\mu\nu \beta}(x,y)
\,,
\end{split}
\ee
and similarly for higher-rank forms if present.
For $\Gseven$, the two-form $\Ab_{\mu\nu\alpha}$ is adjoint-valued, reflecting our use of the index $\alpha$ here, but for other exceptional groups the representations involved will be different.
Note that for $\Gseven$ one also has an additional `constrained compensator' two-form $\Ab_{\mu\nu M}$, see \cite{Hohm:2013uia}, for which the appropriate ansatz in this case is \cite{Hohm:2014qga} $\Ab_{\mu\nu M}(x,y) = - 2 \Delta^2 (y) E^P{}_B(x,y) \partial_M E_P{}^A(x,y) t^\alpha{}_A{}^B \Bflat_{\mu\nu \alpha} (x,y) + \Delta(y) E_M{}^A \Bflat_{\mu\nu A}(x,y)$.
It can then easily be checked that
\be
\mathcal{D}_\mu g_{\nu\rho} = \Delta^2 \Df_\mu \bar g_{\nu\rho} \,,\quad
\mathcal{D}_\mu \gM_{MN} = E_M{}^A E_N{}^B \Df_\mu \Mflat_{AB} \,,\quad
\mathcal{F}_{\mu\nu}{}^M = \Delta E^M{}_A \Fflat_{\mu\nu}{}^A \,,
\ee
where the covariant derivatives are 
\begin{align} \Df_\mu \bar g_{\nu\rho} & = \partial_\mu \bar g_{\nu \rho} - 2 \Aflat_\mu{}^A \theta_A \bar g_{\nu\rho} - \bar{\mathcal{L}}_{\Aflat_\mu} \bar g_{\nu\rho} \,,\\
\Df_\mu \Mflat_{AB} & =  
2\partial_\mu E_P{}^C E^P{}_{(A} \Mflat_{B)C}
- 2 \Aflat_\mu{}^C \left( \Theta_{C(A}{}^D \Mflat_{B)D} + \alpha \tfrac{n-2}{n-1} \mathbb{P}^D{}_{(A|}{}^E{}_C \theta_E \Mflat_{B)D}\right)
- \bar{\mathcal{L}}_{\Aflat_\mu}  \Mflat_{AB}\,,
\end{align}
where $\bar{\mathcal{L}}_{\Aflat_\mu}$ takes the same form as the usual generalised Lie derivative expressed in terms of the indices $A,B,\dots$, partial derivatives $\hat \partial_A$, and the barred fields appearing in the expansion in terms of $E_M{}^A$.
The field strength of $\Aflat_\mu{}^A$ is, specialising here to the $\Gseven$ case for definiteness:
\be
\begin{split}
\Fflat_{\mu\nu}{}^A & = 
2 E_N{}^A \partial_{[\mu} ( E^N{}_B \Aflat_{\nu]}{}^B ) 
+ X_{BC}{}^A \Aflat_{[\mu}{}^B \Aflat_{\nu]}{}^C 
+ ( \Theta^{A \alpha}  - 16 t^{\alpha AB } \theta_B ) \Bflat_{\mu\nu \alpha} 
\\ & \quad
- \bar{\mathcal{L}}_{\Aflat_{[\mu}} \Aflat_{\nu]}{}^A 
-12 t^{\beta AB} \hat \partial_B \Bflat_{\mu\nu \beta} - \tfrac12 \Omega^{AB} \Bflat_{\mu\nu B}
\,.
\end{split}
\ee
For other groups, the differences will be in the representation structure and coefficients of the term involving the two-form $\Bflat_{\mu\nu \alpha}$, while the compensator two-form $\Bflat_{\mu\nu A}$ will be absent.
\begin{comment}
Taking further derivatives we define the action of the derivative $\Df_\mu$ on $\Fflat_{\mu\nu}{}^A$ and $\Df_\mu \Mflat_{AB}$, via:
\be
\begin{split} 
\Df_\mu \Fflat_{\nu\rho}{}^A & = E^A{}_N \partial_\mu ( E^N{}_B \Fflat_{\nu\rho}{}^B) + \Aflat_\mu{}^C X_{CB}{}^A \Fflat_{\nu\rho}{}^B \,,\\%+ E^A{}_N \partial_\mu E^N{}_B F_{\nu\rho}{}^B \,,\\
\Df_\mu (\Df_\nu \Mflat_{AB})& = 
E^M{}_A E^N{}_B \partial_\mu ( E^C{}_M E^D{}_N \Df_\nu \Mflat_{CD} ) 
%+\partial_\mu (\Df_\nu \Mflat_{AB} ) + 2 \partial_\mu E^C{}_P E^P{}_{(A} \Df_\nu \Mflat_{B)C} 
- \Aflat_\mu{}^C \hat \partial_C (\Df_\nu \Mflat_{AB} ) 
\\ & \qquad
- 2 \Aflat_\mu{}^C \left(
\Theta_{C(A}{}^D + \alpha \tfrac{n-2}{n-1} \mathbb{P}^D{}_{(A|}{}^E{}_{C} \theta_E
\right) \Df_\nu \Mflat_{B)D}\,.
\end{split}
\ee
\end{comment}
Then, with external indices raised with $\bar g^{\mu\nu}$, we have the following.

\noindent \emph{Einstein equation.} The Einstein equation \eqref{ExcEin} is:
\be
\begin{split}
0 & = 
R_{\mu \nu}(\bar g) - \tfrac{1}{2} \bar g_{\mu\nu} \left( R(\bar g)+ \tfrac{1}{4 \alpha} \bar g^{\rho\sigma} \Df_\rho \Mflat_{AB} {D}_\sigma \Mflat^{AB} - \tfrac{c_A}{4} \Mflat_{AB} \Fflat_{\rho \sigma}{}^A \Fflat^{\rho \sigma B} - V_{\text{eff}}\right) 
\\ & \qquad
+ \tfrac{1}{4 \alpha} \Df_\mu \Mflat_{AB} \Df_\nu \Mflat^{AB} 
- \tfrac{c_A}{2 } \Mflat_{AB} \Fflat_{\mu \rho}{}^A \Fflat_{\nu}{}^\rho{}^B 
\\ & \qquad
- \Mflat^{AB} \big(
\tfrac14 \hat\partial_A \ln |\bar g | \,\hat\partial_B \bar g_{\mu\nu} 
+ \tfrac12 \bar g_{\mu\rho} \hat\partial_A \bar g^{\rho \sigma} \hat\partial_B \bar g_{\sigma \nu} 
+ \tfrac12 \hat\partial_A \hat\partial_B \bar g_{\mu\nu} 
+ \tfrac{n-2}{2} \theta_A \hat\partial_B \bar g_{\mu\nu} 
\big) 
 \,,
\end{split} 
\label{ExcEin_Ansatzed} 
\ee 	
where $V_{\text{eff}}$ is defined in \eqref{Veff}.

\noindent \emph{Scalar equation.} The scalar equation \eqref{scalarEom} is:
\be
\begin{split} 
0 & = 
 \Mflat_{E(A} \mathbb{P}^{E}{}_{B)}{}^C{}_F \Mflat^{FD} \big(
 -\tfrac{1}{2\alpha}  |\bar g|^{-1/2} \Df_\mu ( |\bar g|^{1/2} \Df^\mu \Mflat_{CD} ) 
+ \tfrac{1}{2\alpha} \Mflat^{GH} \Df_\mu \Mflat_{CG} \Df^\mu \Mflat_{DH} 
\\ & \qquad\qquad \qquad\qquad\quad
+ \tfrac{c_A}{2} \Mflat_{CG} \Mflat_{DH} \Fflat_{\mu\nu}{}^G \Fflat^{\mu\nu H} 
\big) 
+ V_{AB}  \,,
\label{ScalarEom_Ansatzed}
\end{split} 
\ee
where $V_{AB}$ is defined in \eqref{VAB}.

\noindent \emph{One-form equation.} The one-form equation \eqref{Aeom} is:
\begin{align}
\notag 
0 & = |\bar g|^{-1/2} \Df_\nu ( |\bar g|^{1/2} \Mflat_{AB} \Fflat^{\nu\mu B} ) 
+ \tfrac{1}{\alpha} \Mflat^{CD} \Df^\mu \Mflat_{BD} \left( 
\Theta_{AC}{}^B - \tfrac{\alpha (n-2)^2}{n-1} \delta_A^B \theta_C 
\right)
\\ & \quad
\label{OneformEom_Ansatzed}
- |\bar g|^{-1/2} \Mflat^{BC} \hat \partial_C( |\bar g|^{1/2} \Df^{\mu} \Mflat_{AB} )  +\bar g^{\mu\nu} \hat \partial_A \bar{D}_\nu \ln |\bar g| + \hat \partial_A \bar{D}_\nu \bar g^{\mu\nu} 
\\ \notag &  \quad
 - \tfrac{1}{2} \bar g^{\mu\lambda} \hat \partial_A \bar g^{\nu\rho} \bar D_\lambda \bar g_{\nu\rho} 
+\bar g^{\mu\lambda} \hat \partial_A \bar g^{\nu\rho} \bar{D}_\nu \bar g_{\rho\lambda} 
+\tfrac{1}{2} \hat \partial_A \bar g^{\mu\nu} \bar{D}_\nu\ln |\bar g|
 \,.
\end{align} 
For the $\Gseven$ case, this is the equation that results after applying the self-duality condition to combine the terms resulting from the kinetic and Chern-Simons variations. 
In this case we also have the self-duality equation which takes an identical form to \eqref{SelfDuality}, namely
\be
\Fflat_{\mu\nu}{}^A = - \tfrac12 |\bar g|^{1/2} \epsilon_{\mu\nu\rho\sigma} \Omega^{AB} \Mflat_{BC} \Fflat^{\rho\sigma C} \,.
\label{SelfDualityFlat}
\ee
The above ansatz can then be adapted to cover the cases of interest in this paper.

\subsection{Lower-dimensional equations of motion for finite consistent truncations}
\label{EomForNonMaximalTruncations}

A special case of the above construction leads to equations of motion which are those of a conventional finite consistent truncation to a lower-dimensional theory.
In this case we have
\be
E^M{}_A(x,y) =  U^M{}_B(y) (\mathcal{V}^{-1})_A{}^B(x) \,, \quad
 U^M{}_A (y) \equiv \mathring{U}^M{}_B(y) (L^{-1})_A{}^B(y) \,,
\label{EforCT}
\ee
and define the scalar matrix, $M_{AB}$, of the consistent truncation via 
\be
M_{AB}(x) = \mathcal{V}_A{}^C(x) \Mflat_{CD} \mathcal{V}_B{}^D(x)\,,
\label{scalarMxCT}
\ee
while $\bar g_{\mu\nu}=\bar g_{\mu\nu}(x)$ and the gauge fields of the previous section are expanded as\footnote{In addition now $\Bflat_{\mu\nu A} = 0$ for the additional compensator two-form.}
\be
\Aflat_\mu{}^A = \mathcal{V}_B{}^A K_{\ivec }{}^B A_\mu{}^{\ivec }(x) \,,\quad
\Bflat_{\mu\nu\alpha} = \mathcal{V}_\alpha{}^\beta J^{\aadj }{}_\beta B_{\mu\nu \aadj }(x) \,.
\ee
These definitions ensure firstly that we can `factor out' the scalar matrix $\mathcal{V}_B{}^A$ within the equations of motion as previously formulated.

Next the expansion of the gauge fields singles out only the components $X_{IA}{}^B$ defined in \eqref{singlettorsion} of the generalised fluxes $X_{AB}{}^C$, which are now defined in terms of $U^M{}_A$ from \eqref{EforCT} and $\Delta$.
We assume these components $X_{IA}{}^B$ are constant.
Acting on the external metric and the scalars we will thus have in particular the following covariant derivatives: 
\be
\begin{split} D_\mu \bar g_{\nu\rho} & = \partial_\mu \bar g_{\nu \rho} - 2 A_\mu{}^{\ivec } \theta_{\ivec } \bar g_{\nu\rho} \,,\\
D_\mu M_{AB} & = 
\partial_\mu M_{AB}
-  2 A_\mu{}^{\ivec } X_{\ivec  (A}{}^C M_{B)C} - 2 A_\mu{}^{\ivec } \theta_{\ivec } M_{AB} 
\,,
\end{split}
\label{ConvariantDerivative_CT}
\ee
as well as
\be
\begin{split} 
D_\mu F_{\nu\rho}{}^{\ivec } & = \partial_\mu  F_{\nu\rho}{}^{\ivec } + A_\mu{}^{\jvec } X_{\jvec \kvec }{}^{\ivec } F_{\nu\rho}{}^{\kvec } \,,\\%+ E^A{}_N \partial_\mu E^N{}_B F_{\nu\rho}{}^B \,,\\
D_\mu (D_\nu M_{AB})& = 
\partial_\mu (D_\nu M_{AB} ) 
- 2 A_\mu{}^{\ivec } X_{\ivec  (A}{}^C D_\nu M_{B)C} 
-2 A_{\mu}{}^{\ivec }\theta_{\ivec } D_\nu M_{AB} \,.
\end{split}
\label{ConvariantDerivativeGauge_CT}
\ee

\noindent \emph{Einstein equation.} 
Working through the above definitions and logic, the Einstein equation \eqref{ExcEin_Ansatzed} becomes:
\begin{align}
0 & = \notag
R_{\mu \nu}(\bar g) - \tfrac{1}{2} \bar g_{\mu\nu} \left( R(\bar g)+ \tfrac{1}{4 \alpha} \bar g^{\rho\sigma} D_\rho M_{AB} {D}_\sigma M^{AB} - \tfrac{c_A}{4} M_{\ivec \jvec } F_{\rho \sigma}{}^{\ivec } F^{\rho \sigma \jvec } - V_{\text{eff}}\right) 
\\ & \qquad
+ \tfrac{1}{4 \alpha} D_\mu M_{AB} D_\nu M^{AB} 
- \tfrac{c_A}{2 } M_{\ivec \jvec } F_{\mu \rho}{}^{\ivec } F_{\nu}{}^\rho{}^{\jvec }
 \,,
\label{ExcEin_CT} 
\\ \notag
V_{\text{eff}} & \equiv     \tfrac{1}{2\alpha \kappa}(  M^{AB} M^{CD} M_{EF} \Theta_{AC}{}^E \Theta_{BD}{}^F 
 +\kappa M^{AB} \Theta_{AC}{}^D \Theta_{BD}{}^C ) %\notag
 \\  & \qquad  +  \tfrac{(n-2)^3}{n-1}  M^{AB} \theta_A \theta_B
  + 2(n-2)M^{AB} \tilde{\partial}_A \theta_B  \,.
\label{Veff_CT}
\end{align}
Note that the scalar matrix $M_{AB}$ encodes a set of scalar fields $\phi^{\mathcal{I}}$, and 
we can always write their contributions to the Einstein equation as
\begin{equation}
\tfrac{1}{4 \alpha} {D}_\mu M_{AB} {D}_\nu M^{AB}= \mathcal{G}_{\mathcal{I} \mathcal{J}}  D_\mu \phi^\mathcal{I}  D_\nu \phi^\mathcal{J}\,,
\end{equation}
where $\mathcal{G}_{\mathcal{I} \mathcal{J}}$ is the scalar metric of the non-linear sigma model given by $\tfrac{1}{4\alpha}{\partial}_\mu M_{AB} {\partial}^\mu M^{AB}$.
The covariant derivatives $D_\mu$ express the gaugings arising from the components of the intrinsic torsion singled out by the vector fields.
Here we have not attempted to simplify the form of $V_{\text{eff}}$ and left it written in terms of the $Y$-dependent $\Theta$ and $\theta$ of \eqref{IntrinsicTorsionComponentsForNonMaximalTruncations}.
The final term includes a derivative $\tilde \partial_A \equiv \Delta  U^M{}_A \partial_M$.
To have a standard dimensional reduction, we require that $V_{\text{eff}}$ does not depend on the internal coordinates.

\noindent \emph{Scalar equation.} The scalar equation \eqref{ScalarEom_Ansatzed} becomes:
\begin{align} 
0 & = \notag
 M_{I(A} \mathbb{P}^{I}{}_{B)}{}^C{}_E M^{ED} \big(
 -\tfrac{1}{2\alpha}  |\bar g|^{-1/2} D_\mu ( |\bar g|^{1/2} D^\mu M_{CD} ) 
+ \tfrac{1}{2\alpha} M^{FG} D_\mu M_{CF} D^\mu M_{DG} 
\\ & \qquad\qquad \qquad\qquad\quad
+ \tfrac{c_A}{2} M_{CF} M_{DG} F_{\mu\nu}{}^F F^{\mu\nu G} 
\big) 
+ V_{AB}  \,,
\label{ScalarEom_CT}
\\ \notag
- V_{AB} &  = 
M^{CD} M_{E(A|} \tilde{\partial}_C \Theta_{D|B)}{}^E 
+  \alpha\tfrac{n(n-2)}{(n-1)}  M_{E(A} \mathbb{P}^E{}_{B)}{}^C{}_F M^{DF} \tilde{\partial}_C \theta_D 
\\ & \quad 
+ (n-2)   \theta_C M_{E(A} M^{CD} \Theta_{|D|B)}{}^E
\\ \notag & \quad+\tfrac{1}{2 \kappa} M_{A'(A} \mathbb{P}^{A'}{}_{B)}{}^C{}_{D'} M^{DD'}
   \Big( M^{EF} M_{GH} ( \Theta_{CE}{}^G \Theta_{DF}{}^H+ \Theta_{EC}{}^G \Theta_{FD}{}^H)  
 \\  & \qquad \quad\quad \quad\quad \quad \quad \quad \quad \quad \quad \quad \quad 
%\\ &  \qquad  \quad\quad \quad \quad \quad \quad \quad \quad \quad \quad 
+ \kappa \Theta_{CE}{}^F \Theta_{DF}{}^E- M^{EF} M^{GH} M_{CG'} M_{DH'} \Theta_{EG}{}^{G'} \Theta_{FH}{}^{H'} 
\Big)\,.\notag
\label{VAB_CT} 
\end{align}
It is clear that all the terms in this equation other than $V_{AB}$ are automatically independent of internal coordinates.

\noindent \emph{One-form equation.} 
Using the above definitions, and the fact that $M^{CD} D^\mu M_{BD} $ is  $\mathcal{C}_{\Edd}\left( H \right)$-valued, the equations of motion of the one-forms can be shown to be:
\begin{equation}
0  = |\bar g|^{-1/2} D_\nu ( |\bar g|^{1/2} M_{\ivec \jvec } F^{\nu\mu \jvec } ) 
+ \tfrac{1}{\alpha} M^{CD} D^\mu M_{BD} \left( 
%\Theta_{\ivec C}{}^B - \tfrac{\alpha (n-2)^2}{n-1} K_{\ivec }{}^B \theta_C 
X_{\ivec C}{}^B - \alpha (n-2) K_{\ivec }{}^B \theta_C 
\right)\,.
\label{OneFormEquationForHstructReduced}
\end{equation}
This assumes that the consistency conditions discussed between \eqref{Fpartway} and \eqref{FieldStrengthsTruncation} holds.
For $\Gseven$, we should also take into account the self-duality condition \eqref{SelfDualityFlat}.
As $\Omega^{AB}$ is a $H$-invariant tensor, the self-duality condition is projected by a similar argument to that of $M_{AB}$, leading to the reduced self-duality condition:
\begin{equation}
F^{\ivec }=-\Omega^{\ivec \jvec } M_{\jvec \kvec } * F^{ \kvec }\,.
\label{SelfDuality_CT_Recast}
\end{equation}

\section{Details for $\Gseven$} 
\label{app_7rep} 

The fundamental of $\Gseven$ is $R_1 =\mathbf{56}$ and the adjoint is $R_2 = \mathbf{133}$.
Let $M,N,\dots$ denote fundamental indices, and $\alpha, \beta,\dots$ denote adjoint indices.
Denote the generators in the fundamental by $(t_\alpha)_M{}^N$.
In ExFT, we make use of the adjoint projector defined as \cite{Hohm:2013uia}
\be
\mathbb{P}^K{}_M{}^L{}_N = ( t_\alpha )_M{}^K (t^\alpha){}_N{}^L 
= \tfrac{1}{24} \delta^K_M \delta^L_N + \tfrac{1}{12} \delta^L_M \delta^K_N - \tfrac{1}{24} \Omega_{MN} \Omega^{KL} + (t_\alpha)_{MN} (t^\alpha)^{KL} \,,
\label{adjprojE7}
\ee
where $(t_\alpha)_{MN}$ is symmetric.
Fundamental indices are raised and lowered with the antisymmetric invariant $\Omega_{MN}$ such that
$V^M = \Omega^{MN} V_N$, $V_M=V^N\Omega_{NM}$, $\Omega^{MK} \Omega_{NK} = \delta^M_N$,
while adjoint indices are raised and lowered using the Killing form, $\kappa_{\alpha\beta} =(t_\alpha)_M{}^N (t_\beta)_N{}^M$.
In defining the embedding tensor, we use the following projector \cite{deWit:2002vt}
\be
P_{\bf 912}{}_A{}^\alpha,{}^B{}_\beta 
= \tfrac17 ( \delta_A^B \delta^\alpha_\beta - \alpha t^\alpha{}_E{}^B t_\beta{}_A{}^E + \tfrac{\alpha}{n-1} t^\alpha{}_A{}^E t_\beta{}_E{}^B )\,,
\ee
such that in fundamental indices
\be
P_{\bf 912}{}_{AB}{}^C,{}^{DE}{}_F 
= \tfrac17 ( \delta_A^D \mathbb{P}^C{}_B{}^E{}_F- \alpha \mathbb{P}^C{}_B{}^D{}_G \mathbb{P}^E{}_F{}^G{}_A + \tfrac{\alpha}{n-1} \mathbb{P}^C{}_B{}^G{}_A \mathbb{P}^D{}_G{}^E{}_F )\,,
\ee
Acting on $\Omega_{\bullet E}{}^F$ which is already adjoint projected we can replace $\mathbb{P}^E{}_F{}^G{}_H$ with $\delta^E_H\delta_F^G$.
Hence
\be
P_{\bf 912}{}_{AB}{}^C,{}^{DE}{}_F  \Omega_{D E}{}^F 
= \tfrac17 ( \delta_A^D \delta^C{}_F \delta^B{}_E - \alpha \mathbb{P}^C{}_B{}^D{}_F\delta^E_A  + \tfrac{\alpha}{n-1} \mathbb{P}^C{}_B{}^E{}_A \delta^D_F)
\Omega_{D E}{}^F 
= \tfrac17 \Theta_{AB}{}^C \,.
\ee

\paragraph{$\mathrm{SL}(8)$ basis} 

Let $\as,\bs=1,\dots,8$ denote fundamental indices of $\mathrm{SL}(8)$.
The fundamental of $\Gseven$ decomposes as $V^M = (V^{\as\bs}, V_{\as \bs})$, where $V^{\as\bs}$ and $V_{\as \bs}$ are antisymmetric.
We use the contraction convention $V^M W_M = \tfrac12 (V^{\as\bs} W_{\as \bs} + V_{\as \bs} W^{\as \bs} )$.
The adjoint decomposes as $t_\alpha = (t_\as{}^\bs, t_{\as\bs\cs\ds} )$
and the non-zero components of the generators are, following the conventions of \cite{Guarino:2015qaa}:
\be
\begin{split}
(t_\as{}^\bs)_{\cs \ds}{}^{\es \fs} & = 4 ( \delta^{\bs}_{[\cs} \delta^{\es\fs}_{\ds]\as} + \tfrac18 \delta^{\bs}_{\as} \delta^{\es\fs}_{\cs \ds} )
= - (t_\as{}^\bs)^{\es \fs}{}_{\cs \ds} \,,\\
(t_{\as \bs \cs \ds})^{\es \fs \gs \hs} &  = 2 \delta^{\es \fs \gs \hs}_{\as \bs \cs \ds} \,,\quad
(t_{\as \bs \cs \ds}){}_{\es \fs \gs \hs} = \tfrac{1}{12} \epsilon_{\as \bs \cs \ds \es \fs \gs \hs} \,.
\end{split} 
\ee
Then the Killing form has non-vanishing components
\be
\kappa_{\as}{}^{\bs}{}_{\cs}{}^{\ds} = 12 ( \delta^{\bs}_{\cs} \delta^{\ds}_{\as} - \tfrac18 \delta^\bs_\as \delta^\ds_\cs) \,,\quad
\kappa_{\as \bs \cs \ds ,\es \fs \gs \hs} =  \tfrac{1}{12} \epsilon_{\as \bs \cs \ds \es \fs \gs \hs} \,.
\ee
%with inverse
%\be
%(\kappa^{-1})_{\as}{}^{\bs}{}_{\cs}{}^{\ds} = \tfrac{1}{12} ( \delta^{\bs}_{\cs} \delta^{\ds}_{\as} - %\tfrac18 \delta^\bs_\as \delta^\ds_\cs) \,,\quad
%(\kappa^{-1})^{\as \bs \cs \ds ,\es \fs \gs \hs} =  12 \epsilon^{\as \bs \cs \ds \es \fs \gs \hs} \,,
%\ee
We can write the adjoint projector used in ExFT as:
\be
\mathbb{P}^K{}_M{}^L{}_N = \tfrac{1}{12} ( t_{\es}{}^{\fs})_M{}^K (t_{\fs}{}^{\es}){}_N{}^L
+ \tfrac12 \epsilon^{\es_1 \dots \es_4 \fs_1 \dots \fs_4} ( t_{\es_1 \dots \es_4})_M{}^K ( t_{\fs_1 \dots \fs_4})_N{}^L \,.
\ee

\paragraph{$\mathrm{SO}(8)$ branching}
The $\mathrm{SO}(8)$ branching is performed by identifying fundamental $\mathrm{SL}(8)$ indices with $\mathbf{8}_v$ indices. One can now raise and lower indices with the $\mathrm{SO}(8)$ invariant tensor $\delta_{\as \bs}$. The adjoint of $E_{7(7)}$ branches as: $\mathbf{133} \rightarrow \mathbf{28} + \mathbf{35}_v + \mathbf{35}_c + \mathbf{35}_s$, with the corresponding identification of the generators as follows:
\begin{equation}
R_{\as \bs}= 2 t_{[\as}{}^\cs \delta_{\bs]\cs}\,,\quad
S_{\as \bs}= 2 t_{(\as}{}^\cs \delta_{\bs)\cs}\,,\quad
t^\pm_{\as\bs\cs\ds}=\tfrac{1}{24} \left( t_{\as\bs\cs\ds} \pm \epsilon_{\as\bs\cs\ds \es \fs \gs \hs} t^{\es \fs \gs \hs} \right)\,.
\label{Rstpm}
\end{equation}

\section{Details for the case $\Gsym = \mathrm{SU}(4)$}
\label{appSU4}

Here we give some details corresponding to the example worked out in section \ref{sec:ExampleSU4}.

\paragraph{The coset $\mathrm{SU}(4)/\mathrm{SU}(3)$}
We construct the coset $\mathrm{SU}(4)/\mathrm{SU}(3)$ following \cite{Bertini:2005rc}. Consider the Gell-Mann-like matrices for $\mathrm{SU}(4)$:
\begin{equation}
\small
\begin{matrix}
a_{(1)} = 
\begin{pmatrix}
 0 & i & 0 & 0 \\
 i & 0 & 0 & 0 \\
 0 & 0 & 0 & 0 \\
 0 & 0 & 0 & 0 \\
\end{pmatrix}
,
&a_{(2)}=
\begin{pmatrix}
 0 & 1 & 0 & 0 \\
 -1 & 0 & 0 & 0 \\
 0 & 0 & 0 & 0 \\
 0 & 0 & 0 & 0 \\
\end{pmatrix}
,
&a_{(3)}=
\begin{pmatrix}
 i & 0 & 0 & 0 \\
 0 & -i & 0 & 0 \\
 0 & 0 & 0 & 0 \\
 0 & 0 & 0 & 0 \\
\end{pmatrix}
,\\\\
a_{(4)}=
\begin{pmatrix}
 0 & 0 & i & 0 \\
 0 & 0 & 0 & 0 \\
 i & 0 & 0 & 0 \\
 0 & 0 & 0 & 0 \\
\end{pmatrix}
,
&a_{(5)}=
\begin{pmatrix}
 0 & 0 & 1 & 0 \\
 0 & 0 & 0 & 0 \\
 -1 & 0 & 0 & 0 \\
 0 & 0 & 0 & 0 \\
\end{pmatrix}
,
&a_{(6)}=
\begin{pmatrix}
 0 & 0 & 0 & 0 \\
 0 & 0 & i & 0 \\
 0 & i & 0 & 0 \\
 0 & 0 & 0 & 0 \\
\end{pmatrix}
,\\\\
a_{(7)}=
\begin{pmatrix}
 0 & 0 & 0 & 0 \\
 0 & 0 & 1 & 0 \\
 0 & -1 & 0 & 0 \\
 0 & 0 & 0 & 0 \\
\end{pmatrix}
,
&a_{(8)}=
\begin{pmatrix}
 \frac{i}{\sqrt{3}} & 0 & 0 & 0 \\
 0 & \frac{i}{\sqrt{3}} & 0 & 0 \\
 0 & 0 & -\frac{2 i}{\sqrt{3}} & 0 \\
 0 & 0 & 0 & 0 \\
\end{pmatrix}
,
&a_{(9)}=
\begin{pmatrix}
 0 & 0 & 0 & i \\
 0 & 0 & 0 & 0 \\
 0 & 0 & 0 & 0 \\
 i & 0 & 0 & 0 \\
\end{pmatrix}
,\\\\
a_{(10)}=
\begin{pmatrix}
 0 & 0 & 0 & 1 \\
 0 & 0 & 0 & 0 \\
 0 & 0 & 0 & 0 \\
 -1 & 0 & 0 & 0 \\
\end{pmatrix}
,
&a_{(11)}=
\begin{pmatrix}
 0 & 0 & 0 & 0 \\
 0 & 0 & 0 & i \\
 0 & 0 & 0 & 0 \\
 0 & i & 0 & 0 \\
\end{pmatrix}
,
&a_{(12)}=
\begin{pmatrix}
 0 & 0 & 0 & 0 \\
 0 & 0 & 0 & 1 \\
 0 & 0 & 0 & 0 \\
 0 & -1 & 0 & 0 \\
\end{pmatrix}
,\\\\
a_{(13)}=
\begin{pmatrix}
 0 & 0 & 0 & 0 \\
 0 & 0 & 0 & 0 \\
 0 & 0 & 0 & i \\
 0 & 0 & i & 0 \\
\end{pmatrix}
,
&a_{(14)}=
\begin{pmatrix}
 0 & 0 & 0 & 0 \\
 0 & 0 & 0 & 0 \\
 0 & 0 & 0 & 1 \\
 0 & 0 & -1 & 0 \\
\end{pmatrix}
,
&a_{(15)}=
\begin{pmatrix}
 \frac{i}{\sqrt{6}} & 0 & 0 & 0 \\
 0 & \frac{i}{\sqrt{6}} & 0 & 0 \\
 0 & 0 & \frac{i}{\sqrt{6}} & 0 \\
 0 & 0 & 0 & -i \sqrt{\frac{3}{2}} \\
\end{pmatrix}
\,.
\end{matrix}
\label{SU4cGenerators8v}
\end{equation}
We can view these matrices as acting on the complex space $\mathbb{C}^4$ with coordinates $\vec{z} = ( z_1,z_2,z_3,z_4)$. 
We can identify $\mathbb{C}^4 \cong \mathbb{R}^8$ and define real coordinates $\vec{x} = (x^{\as})$ such that 

\begin{equation}
\vec{z}=\left( \begin{matrix}
z_1\\
z_2\\
z_3\\
z_4
\end{matrix}\right)=
\left( \begin{matrix}
\frac{1}{\sqrt{2}}\left(x^1+ix^2\right)\\
\frac{1}{\sqrt{2}}\left(x^3+ix^4\right)\\
\frac{1}{\sqrt{2}}\left(x^5+ix^6\right)\\
\frac{1}{\sqrt{2}}\left(x^7+ix^8\right)
\end{matrix}\right)\,,\quad
\begin{pmatrix}
\vec{z}\\
\vec{\bar{z}}
\end{pmatrix} = U \vec{x}\,,
\end{equation}
which can be seen just as a unitary rotation of the $\mathbf{8}_v$ representation of $\mathrm{SO}(8)$.
The representation of the Gell-Mann matrices on the $\mathbf{8}_v$ of $\mathrm{SO}(8)$ is accordingly given by:
\begin{equation}
a_{(\mu)}^{(8_v)} = U^{-1} \left(
\begin{matrix}
&a_{(\mu)}   &\mathbf{0}\\
&\mathbf{0}               &\bar{a}_{(\mu)}
\end{matrix}
\right) U \,.
\end{equation}
These matrices provide a (real) representation of $\mathrm{SU}(4)$ acting on the $8_v$ of $\mathrm{SO}(8)$. 
In terms of the above coordinates, the $\mathrm{SU}(4)$ invariant tensors $\{\Omega, J\}$ can be defined via:
\begin{equation}
\begin{split}
\Omega & = (\dd x^1 + i\dd x^2)\wedge  (\dd x^3 + i\dd x^4) \wedge  (\dd x^5 + i\dd x^6) \wedge  (\dd x^7 + i\dd x^8) \,, \\
 J & =\dd x^1 \wedge \dd x^2 + \dd x^3 \wedge \dd x^4+\dd x^5 \wedge \dd x^6+\dd x^7 \wedge \dd x^8\,.
\end{split}
\label{OmegaJ}
\end{equation}
Now we construct the $\mathrm{SU}(4)/\mathrm{SU}(3)$ coset element $L(y)$ acting on the $\mathbf{8}_v$ of $\mathrm{SO}(8)$ as
Following \cite{Bertini:2005rc}, it can be obtained by exponentiating in terms of Euler angles as:
\begin{equation}
L(y)^{(8_v)}=e^{\theta_1 a_{(3)}^{(8_v)}} e^{\phi_1 a_{(2)}^{(8_v)}} e^{\theta_2 a_{(3)}^{(8_v)}} e^{\phi_2 a_{(5)}^{(8_v)}} e^{\frac{1}{\sqrt{3}}\theta_3 a_{(8)}^{(8_v)}} e^{\phi_3 a_{(10)}^{(8_v)}} e^{ \frac{1}{\sqrt{3}} \phi_* a_{(15)}^{(8_v)}} \,.
\label{SU4cL(y)8v}
\end{equation}
The embedding coordinates of $S^7$ in terms of this particular choice of angular coordinates are:
\begin{equation}
\mu^\as=(L(y)^{(8_v)}\,  )_\bs{}^\as (v_0)^\bs \,, \quad v_0=(0\,,0\,,0\,,0\,,0\,,0\,,0\,,1) \,.
\end{equation}
The 56-dimensional representation of $L(y)$ is then obtained just by the standard embedding of $\mathrm{SO}(8)$ into $\mathrm{SL}(8)$ inside $\Gseven$:
\begin{equation}
L(y)  = \begin{pmatrix} 
 2 (L(y)^{(8_v)})_\cs{}^{[\as} (L(y)^{(8_v)})_\ds{}^{\bs]} & 0 \\
 0 & 2 (L^{-1}(y)^{(8_v)})_{[\as}{}^\cs(L^{-1}(y)^{(8_v)})_{\bs]}{}^\ds
 \end{pmatrix}\,.
\label{SU4cL(y)56from8v}
\end{equation}

\paragraph{The $\mathrm{SU}(3)$-structure}

The $\mathrm{SU}(3)$-structure is given by the generators of $\mathrm{SU}(3)$ embedded into $\Gseven$ via
\begin{equation}
( a_{(\mu)}^{(8_v)} )^\as{}_\bs\, t_\as{}^\bs\,,
\end{equation}
where $ t_\as{}^\bs$ are the generators of $\mathrm{SL}(8)\subset E_{7(7)}$. These $\mathrm{SU}(3)$ generators coincide with those of \cite{Larios:2019kbw}, so we borrow here their notation and conventions. The branching of the $\mathbf{8}_v$ of $\mathrm{SO}(8)$ under $\mathrm{SU}(3)$ reads $\mathbf{8}_v \rightarrow \mathbf{3} + \bar{\mathbf{3}} + \mathbf{1} +\mathbf{1}$ and leads to the splitting of the index $\as=(i,7,8)$, where $i$ is the $\mathbf{6}= \mathbf{3} + \bar{\mathbf{3}}$ of $\mathrm{SU}(3)$. We can write $\mathcal{C}_{\Gseven}\left(\mathrm{SU}(3)\right) = \mathrm{SL}(2 ,\mathbb{R}) \times \mathrm{SU}(2,1)$ in terms of the $\mathrm{SU}(3)$ invariant tensors:
\be
\begin{split}
\Omega & = (\dd x^1 + i \dd x^2) \wedge (\dd x^3 + i \dd x^4) \wedge (\dd x^5 + i \dd x^6) \,,\\
J& = \dd x^1 \wedge \dd x^2 + \dd x^3 \wedge \dd x^4 + \dd x^5 \wedge \dd x^6\, . 
\end{split}
\ee
The $\mathrm{SL}(2 ,\mathbb{R})$ factor is generated by 
\begin{eqnarray}
H_0 = -\tfrac12 \big(  t_i{}^i - 3 t_{7}{}^{7}- 3 t_8{}^8 \big) \; , \qquad 
E_0 = 6 \, J^{ ij} \,  t_{ij 78} \; , \qquad 
F_0 = \tfrac32 \, J^{ ij} J^{ kh} \,  t_{ijkh} \; , 
\label{SU4cCommutantGeneratorsSL2}
\end{eqnarray}
while the $\mathrm{SU}(2,1)$ factor is generated by: 
\begin{eqnarray} 
& H_1 = - t_7{}^7 + t_8{}^8    \; , \qquad 
H_2 =  J_j{}^i \,  t_i{}^j \; , \nonumber \\
& E_{11} = -\sqrt{2} \; \textrm{Im} \, \Omega^{ ijk} \;  t_{ijk8} \; , \qquad 
E_{12} = -\sqrt{2} \; \textrm{Re} \, \Omega^{ ijk} \;  t_{ijk8} \; , \qquad 
E_{2} = -\sqrt{2} \; \,  t_8{}^7  \; ,  \nonumber \\
& F_{11} = \sqrt{2} \; \textrm{Re} \, \Omega^{ ijk} \;  t_{ijk7} \; , \qquad 
F_{12} = -\sqrt{2} \; \textrm{Im} \, \Omega^{ ijk} \;  t_{ijk7} \; , \qquad 
F_{2} = -\sqrt{2} \; \,  t_7{}^8 \; . 
\label{SU4cCommutantGeneratorsSU(2,1)}
\end{eqnarray}
The indices $i,j$ and $7,8$ can be raised and lowered with the respective 6- and 2-dimensional Euclidean metrics. The coset representative is defined by:
\begin{equation}
\mathcal{V}_A{}^B = (  e^{ -3 U  H_1 } e^{-\frac{1}{\sqrt{2}}\left( a E_2 - \sqrt{3} \chi_1 E_{11} - \sqrt{3} \chi_2 E_{12} \right)}  e^{-\frac{1}{2}  (2U +V) H_0} e^{h E_0 }   )_{{A}}{}^{B}\,.
\label{SU4cCosetParametrization}
\end{equation}
The $\mathrm{SU}(3)$ invariant vectors $K_{\ivec }{}^A$, $\ivec =1,\dots,4$ are defined through the condition
\begin{equation}
( a_{(\mu)}^{(8_v)} )^\as{}_\bs\, (  t_\as{}^\bs )_B{}^A K_{\ivec }{}^B=0,
\end{equation}
Branching the $\mathbf{56}$ of  $\Gseven$ firstly under $\mathrm{SO}(8)$ and secondly under $\mathrm{SU}(3)$: $(^A) \rightarrow (^{[\as\bs]},_{[\as\bs]}) \rightarrow (^{[ij]},^{[i7]},^{[i8]},^{[78]},_{[ij]},_{[i7]},_{[i8]},_{[78]}) $, where $i$ is the $\mathbf{6}= \mathbf{3} + \bar{\mathbf{3}}$ of $\mathrm{SU}(3)$, one can express the four invariant vectors as:
\begin{equation}
\begin{split}
K_{(1)} & = \tfrac{1}{\sqrt{3}} (J^{ij},0,0,0,0,0,0,0)\,,\quad
K_{(2)} = (0,0,0,1,0,0,0,0)\,,\\
K_{(3)} & = \tfrac{1}{\sqrt{3}} (0,0,0,0,J_{ij},0,0,0)\,,\quad
K_{(4)} = (0,0,0,0,0,0,0,1)\,,\\
\end{split}
\label{SU3InvariantVectorsExplicit}
\end{equation}
where $i,j$ indices are rised and lowered with $\delta_{ij}$, and we have made a convenient choice of normalisation.

%%%%%%%%%%%%%%%%%%%%%%%%%%%%%%%%%%%%%%%%%%%%%%%%%%%%%%%%%%%%%%%%%%%%%%%%
\section{Details for the case $\Gsym = \mathrm{SO}(7)$}
\label{AppendixSO7/SO6Coset}

Here we collect some details pertaining to the SO$(7)_v$ example described in section \ref{sec:ExampleSO7}. 

\paragraph{The coset $\mathrm{SO}(7)/ \mathrm{SO}(6)$}

The generators of $\mathrm{SO}(8)$ in the $8_v$ representation are given by:
\begin{equation}
(R_{\as \bs})^\cs{}_\ds= 2 \delta^\cs_{[\as}\delta_{\bs]\ds} \, .
\end{equation}
The $\mathbf{8}_v$ representation of $\mathrm{SO}(8)$ is branched to the $\mathbf{7}+\mathbf{1}$ under $\mathrm{SO}(7)_v$. Therefore, we have the splitting of the fundamental index $\as=(i,8)$. The $\mathrm{SO}(7)_v$ subgroup of $\mathrm{SO}(8)$ is generated by $R_{ij}$. Under $\mathrm{SO}(6)_v$ the index $i$ further branches as $i=(\overline{\imath},7)$, where $\overline{\imath}$ transforms as the $\mathbf{6}$ of $\mathrm{SO}(6)$. The $\mathrm{SO}(6)_v$ generators are given by $R_{\overline{\imath}\, \overline{\jmath}}$. Following \cite{Castellani:1991et}, we obtain the $\mathrm{SO}(7)_v / \mathrm{SO}(6)_v$ coset by means of the exponential:
\begin{equation}
L(\theta)^{(8_v)}= e^{\sum_{\overline{\imath}} \theta_{\overline{\imath}} R_{\overline{\imath} 7}} \,.
\end{equation}
This can be rewritten in terms of $S^6$ embedding coordinates by means of the identifications $\tilde{\mu}_{\overline{\imath}}= \theta _{\overline{\imath}} \frac{\sin \left(\sqrt{\sum_{\overline{\imath}} \theta_{\overline{\imath}} \theta_{\overline{\imath}}} \right) }{\sqrt{\sum_{\overline{\imath}} \theta_{\overline{\imath}} \theta_{\overline{\imath}}}}$, $\tilde{\mu}_7 = 1 - \sum_{\overline{\imath}} \tilde{\mu}_{\overline{\imath}} \tilde{\mu}_{\overline{\imath}} $, leading to:
\begin{equation}
L(\tilde{\mu})^{(8_v)} = 
\begin{pmatrix}
\delta_{\overline{\imath} \,\overline{\jmath}}
 - \tfrac{\tilde\mu_{\overline{\imath}} \tilde \mu_{\overline{\jmath}}}{1+ \tilde \mu_7} & -\tilde \mu_{\overline{\imath}} & 0 \\
  \tilde\mu_{\overline{\jmath}} & \tilde \mu_7 & 0 \\
 0 & 0 &1 
\end{pmatrix} \,.
\label{SO7vL(y)8v}
\end{equation}
The explicit evaluation of the embedding coordinates $\tilde{\mu}_i$ in terms of internal coordinates give us $L(y)$. The embedding coordinates of $S^7$ with this prescription are obtained as:
\begin{equation}
\mu_i = \sin(\theta_7) \tilde{\mu}_i \,,\quad \mu_8 = \cos (\theta_7) \,.
\label{SO7vEmbeddingCoordinates}
\end{equation}
The 56 dimensional representation of $L(y)$ is then obtained using \eqref{SU4cL(y)56from8v}. 

\paragraph{The $\mathrm{SO}(6)$ structure}

The commutant of $\mathrm{SO}(6)_v$ is given by $\mathcal{C}_{\Gseven}\left(\mathrm{SO}(6)_v\right) = \mathrm{SO}(1,1) \times \mathrm{SL}(2)$. The factor $\mathrm{SO}(1,1)$ is generated by:
\begin{equation}
\tilde{H}_0 = \tfrac12 H_0 = -\frac14 \big(  t_i{}^i - 3 \left( t_{7}{}^{7}+t_{8}{}^{8} \right) \big)\,,
\end{equation}
while the factor $\mathrm{SL}(2)$ is generated by:
\begin{equation}
\tilde{H}_1=\tfrac12 H_1 = - \tfrac12 \left( t_7{}^7 - t_8{}^8  \right) \; , \qquad 
\tilde{E}_1= - \frac{1}{\sqrt{2}} E_2 =t_8{}^7\, , \quad
\tilde{F}_1= - \frac{1}{\sqrt{2}} F_2 =t_7{}^8\, .
\end{equation}
We define the coset representative for $\mathcal{V} \in \mathcal{C}_{\Gseven}( \mathrm{SO}(6)_v)$ / $\mathcal{C}_{\mathrm{SU}(8)}({\mathrm{SO}(6)_v})$ by:
\begin{equation}
\mathcal{V}_A{}^B = (  e^{\frac{1}{2}  \phi(\theta_7) \tilde{H}_0} e^{ \tfrac12 \varphi (\theta_7)  \tilde{H}_1 } e^{\chi (\theta_7) E_{1}})_{{A}}{}^{B}\,.
\label{SO7cosetparam}
\end{equation}
The $\mathrm{SO}(6)$ invariant vectors $K_{\ivec }{}^A$, $\ivec =1,2$ are defined through the condition
\begin{equation}
( R_{\overline{\imath}\, \overline{\jmath}} )^\as{}_\bs\, (  t_\as{}^\bs )^A{}_B K_{\ivec }{}^B=0,
\end{equation}
Branching the $\mathbf{56}$ of  $\Gseven$ firstly $\mathrm{SO}(8)$ and secondly under $\mathrm{SO}(6)_v$, so that $(^A) \rightarrow (^{[\as\bs]},_{[\as\bs]}) \rightarrow (^{[\overline{\imath}\, \overline{\jmath}]},^{[\overline{\imath}7]},^{[\overline{\imath}8]},^{[78]},_{[\overline{\imath}\, \overline{\jmath}]},_{[\overline{\imath}7]},_{[\overline{\imath}8]},_{[78]}) $, one can express the two invariant vectors as:
\begin{equation}
K_{(1)} = (0,0,0,1,0,0,0,0)\,,\quad
K_{(2)} = (0,0,0,0,0,0,0,1)\,.\\
\label{SO6InvariantVectorsExplicit}
\end{equation}

\paragraph{Equations for vacuum solutions}

Here we record the equations of motion arising when assuming the gauge fields vanish, the external metric $\bar g_{\mu\nu}(x,y)$ is $y$-independent  and the scalars are independent of four-dimensional external coordinates. 
Suppressing their remaining $\theta_7$ dependence, the equations that arise for the scalars are:
\begin{equation}
\begin{split}
0&=20(- e^{\varphi} (\chi+\cot (\theta_7))^2+ \sinh (\varphi)- \cosh (\varphi))
 +e^{2 (\varphi+\phi)} \left(\chi'+5 \chi^2+6 \chi \cot (\theta_7)-1\right)^2 \\
&\quad +e^{2 \phi} \Big(-4 \varphi' (5 \chi+3 \cot (\theta_7))+\varphi'^2-10 \chi'+50 \chi^2+60 \chi \cot (\theta_7)\\
&\quad \quad \quad \quad \quad \quad \quad+4 \phi''+6 \phi' \left(\phi'+4 \cot (\theta_7)\right)+20 \csc ^2(\theta_7)-26\Big)
\\ & \quad  +25 ( \sinh (2 [\phi-\varphi ])+ \cosh (2 [\phi-\varphi ]) )
\,,
\end{split}
\label{lovely1}
\end{equation}
\begin{equation}
\begin{split}
0&=-80 e^{\varphi} \left(\cot ^2(\theta_7)-4 \chi  (\chi+\cot (\theta_7))\right)+160 e^{3 \varphi } \chi ^2 (\chi +\cot (\theta_7))^2 + 160 e^{-\varphi }
\\ & \quad +2 e^{2 \phi } \Big(60 \chi  \left(\varphi' -3 \cot (\theta_7)\right)-(\varphi')^2+30 \chi' -225 \chi ^2\\
&\quad \quad \quad \quad \quad  -2 \left(\varphi'' +2 \phi'' +23\right)+6 \phi'  \left(2 \cot (\theta_7)-\varphi' \right) -6 (\phi')^2+40 \csc ^2(\theta_7)\Big)\\
&\quad -2 e^{2 (\varphi +\phi )} \Big(60 \chi ^3 \left(6 \cot (\theta_7)-\varphi' \right)-\left(\chi' -1\right)^2+225 \chi ^4\\
&\quad \quad \quad \quad \quad\quad\, -4 \chi  \Big(2 \varphi'  \left(\chi' -1\right)+\chi''+9 \cot (\theta_7) \chi' +3 \left(\chi' -1\right) \phi' -3 \cot (\theta_7)\Big)\\
&\quad \quad \quad \quad \quad \quad\, +\chi ^2 \Big(-2 \varphi'' +(\varphi')^2+4 \phi'' -6 \varphi'  \left(\phi' +12 \cot (\theta_7)\right) \\
&\quad \quad \quad \quad \quad \quad \quad\qquad +6 \phi'  \left(\phi' -2 \cot (\theta_7)\right)+68 \csc ^2(\theta_7)-92\Big)\Big)\\
&\quad -6 \chi ^2 e^{4 \varphi +2 \phi } \left(\chi' +5 \chi ^2+6 \chi  \cot (\theta_7)-1\right)^2-150 e^{2 \phi -2 \varphi }\,,
\end{split}
\label{lovely2} 
\end{equation}
\begin{equation}
\begin{split}
0&=10 e^{3 \phi } \varphi'  -15e^{3 \phi } \left(\phi' +4 \cot (\theta_7)\right)+60 \cot (\theta_7) e^{\varphi +\phi } \left(e^{2 \varphi } \chi  (\chi +\cot (\theta_7))+1\right)\\
&\quad -e^{3 \phi +4 \varphi } \chi  \Big(2 \left(\chi' -1\right)^2 +5 \chi ^3 \left(2 \left(\varphi' +6 \cot (\theta_7)\right)+3 \phi' \right) \\
&\quad \quad \quad \quad \quad \quad  +2 \chi ^2 \Big[6 \cot (\theta_7) \varphi' +10 \chi' +9 \cot (\theta_7) \phi' +33 \csc ^2(\theta_7)-41\Big]\\
&\quad \quad \quad \quad \quad \quad  +\chi  \Big[2 \varphi'  \left(\chi' -1\right)+\chi''+6 \cot (\theta_7) \left(5 \chi' -4\right) +3 \left(\chi' -1\right) \phi' \Big]\Big)\\
&\quad +e^{3 \phi +2 \varphi }\Big(2 \varphi'  \left(\chi' -1\right)+\chi''+6 \cot (\theta_7) \chi' \\
&\quad \quad \quad \quad \qquad +2 \chi  \Big[\varphi'' -10 \chi' +3 \varphi'  \left(\phi' +4 \cot (\theta_7)\right) -9 \cot (\theta_7) \phi' -33 \csc ^2(\theta_7)+41\Big]\\
&\quad \quad \quad \quad \quad \quad  +3 \left(\chi' -1\right) \phi' -30 \chi ^2 \left(\phi' +4 \cot (\theta_7)\right)\Big)\,.
\end{split}
\label{lovely3}
\end{equation}

\bibliography{ChrisBib.bib}

\end{document}